\documentclass{article}
\pdfoutput=1

\usepackage{arxiv}
\usepackage{amssymb}
\usepackage{amsmath}

\usepackage{algorithm}
\usepackage{algpseudocode}
\def\NoNumber#1{{\def\alglinenumber##1{}\State #1}\addtocounter{ALG@line}{-1}}
\usepackage{adjustbox}
\usepackage{amsfonts}
\usepackage{amssymb}
\usepackage{amsbsy}
\usepackage[T1]{fontenc}
\usepackage{textcomp}
\usepackage{graphicx}
\usepackage{lineno}
\usepackage{listings}
\usepackage{multirow}
\usepackage[round]{natbib} 
\usepackage[normalem]{ulem}
\usepackage{color}
\definecolor{ruby}{rgb}{0.6,0,0.3}
\definecolor{lightcoral}{rgb}{0.94117647,0.50196078,0.50196078}
\definecolor{gold1}{rgb}{0.8667,0.8510,0.7647}
\definecolor{gold2}{rgb}{0.7686,0.7412,0.5922}
\definecolor{gold3}{rgb}{0.5804,0.5412,0.3294}
\definecolor{blue1}{rgb}{0.7765,0.8510,0.9451}
\definecolor{blue2}{rgb}{0.5569,0.7059,0.8902}
\definecolor{blue3}{rgb}{0.3333,0.5569,0.8353}
\definecolor{gray1}{rgb}{0.9,0.9,0.9}
\definecolor{gray2}{rgb}{0.8,0.8,0.8}
\definecolor{gray3}{rgb}{0.7,0.7,0.7}

\usepackage{soul}

\usepackage{epstopdf}
\usepackage[caption=false]{subfig}

\usepackage{multirow}
\usepackage{xparse}
\ExplSyntaxOn
\NewDocumentCommand{\smallcaps}{m}
 {
  \tl_set:Nn \l_tmpa_tl { #1 }
  \regex_replace_all:nnN
   { ([0-9]+) } 
   { \c{resizedigit}\cB\{ \1 \cE\} } 
   \l_tmpa_tl
  \textsc{ \tl_use:N \l_tmpa_tl }
 }

\ExplSyntaxOff

\usepackage{tikz}

\usepackage[hidelinks]{hyperref}
\usepackage{url}

\graphicspath{{./}} 
\renewcommand{\vec}[1]{\boldsymbol{#1}}
\newcommand{\mat}[1]{\mathbf{#1}}
\newcommand{\cov}[0]{\text{cov}}
\newcommand{\erf}[0]{\text{erf}}
\newcommand{\sgn}[0]{\text{sgn}}
\newcommand{\diag}[0]{\text{diag}}
\newcommand\defeq{\mathrel{\overset{\makebox[0pt]{\mbox{\normalfont\tiny def}}}{=}}}
\newcommand{\code}[2][black]{{\color{#1}\small\textbf{\texttt{#2}}}} 
\newcommand{\CC}{C\nolinebreak\hspace{-.05em}\raisebox{.4ex}{\tiny\bf +}\nolinebreak\hspace{-.10em}\raisebox{.4ex}{\tiny\bf +}\,} 
\newenvironment{rcases}
  {\left.\begin{aligned}}
  {\end{aligned}\right\rbrace}

\usepackage{fancyhdr}
\fancypagestyle{specialfooter}{%
  \fancyhf{}
  
  \fancyfoot[L]{\footnotesize Cite as: Chlingaryan,\,A., Melkumyan,\,A., Leung,\,R. (2026). IntegralGP: Volumetric estimation of subterranean geochemical properties in mineral deposits by fusing assay data with different spatial supports, \textit{Expert Systems with Applications}, vol.\,298A, article 129429. DOI: \href{https://doi.org/10.1016/j.eswa.2025.129429}{10.1016/j.eswa.2025.129429}}
}

\usepackage[all]{background} 
\usepackage{lipsum}

\SetBgColor{gray}
\SetBgContents{\scriptsize \fbox{\begin{minipage}{100mm}{Article published in \textbf{Expert Systems with Applications}. \href{https://doi.org/10.1016/j.eswa.2025.129429}{DOI: 10.1016/j.eswa.2025.129429}\\ \copyright 2025. This manuscript version is made available under the \href{https://creativecommons.org/licenses/by/4.0/}{CC BY 4.0} license.}\end{minipage}}}
\SetBgPosition{current page.east}
\SetBgVshift{+1.2cm}
\SetBgOpacity{0.75}
\SetBgAngle{90.0}
\SetBgScale{1.25}

\title{\textbf{\Large{IntegralGP: Volumetric estimation of subterranean geochemical properties in mineral deposits by fusing assay data with different spatial supports}}}

\date{\vspace{-5mm}}

\author{
  \textbf{Anna~Chlingaryan$^\dag$, Arman~Melkumyan$^\star$, Raymond~Leung$^\star$}\vspace{2mm} \\
  $\dag$ School of Life and Environmental Sciences, Faculty of Science\\
  $\star$ Rio Tinto Sydney Innovation Hub, Faculty of Engineering\\
  The University of Sydney\\
}

\renewcommand{\overtitle}{}
\renewcommand{\headerleft}{IntegralGP: volumetric estimation of subterranean geochemical properties}
\renewcommand{\headeright}{Chlingaryan, Melkumyan and Leung}
\renewcommand{\undertitle}{\textbf{Accepted article}}

\begin{document}
\maketitle

\begin{abstract}
This article presents an Integral Gaussian Process (IntegralGP) framework for volumetric estimation of subterranean properties in mineral deposits. It provides a unified representation for data with different spatial supports, which enables blasthole geochemical assays to be properly modelled as interval observations rather than points. This approach is shown to improve regression performance and boundary delineation. A core contribution is a description of the mathematical changes to the covariance expressions which allow these benefits to be realised. The gradient and anti-derivatives are obtained to facilitate learning of the kernel hyperparameters. Numerical stability issues are also discussed. To illustrate its application, an IntegralGP data fusion algorithm is described. The objective is to assimilate line-based blasthole assays and update a block model that provides long-range prediction of Fe concentration beneath the drilled bench. Heteroscedastic GP is used to fuse chemically compatible but spatially incongruous data with different resolutions and sample spacings. Domain knowledge embodied in the structure and empirical distribution of the block model must be generally preserved while local inaccuracies are corrected. Using validation measurements within the predicted bench, our experiments demonstrate an improvement in bench-below grade prediction performance. For material classification, IntegralGP fusion reduces the absolute error and model bias in categorical prediction, especially instances where waste blocks are mistakenly classified as high-grade.\\
\end{abstract}

\begin{centering}
\begin{tabular}{p{160mm}}
\multicolumn{1}{c}{\large\textsc{\textbf{Highlights}}\vspace{3mm}}\\
\hspace{10mm}\texttt{\small - Integral GP enables volumetric regression using data with different spatial supports}\\
\hspace{10mm}\texttt{\small - For subterranean attribute estimation, it can improve accuracy \& boundary delineation}\\
\hspace{10mm}\texttt{\small - New covariance expressions are developed while GP learning/inference remain unchanged}\\
\hspace{10mm}\texttt{\small - Heteroscedastic GP fusion is used to improve bench-below grade prediction performance}\\
\hspace{10mm}\texttt{\small - Proposed method reduces model bias and the absolute error in material classification}\\
\end{tabular}
\end{centering}

\begin{tabular}{cp{45mm}} 
&\\
\multirow{1}{*}{\includegraphics[width=0.62\textwidth,trim={0mm 0 0 0},clip]{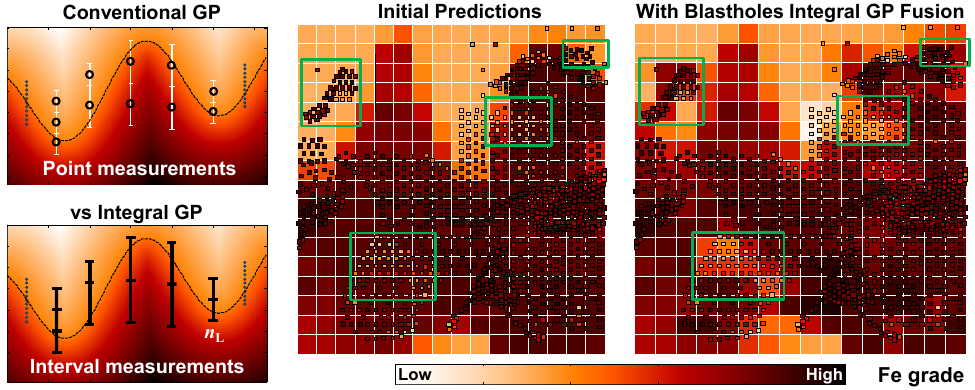}} &\\\\
& {\sffamily\scriptsize \textbf{Keywords}: Heteroscedastic Gaussian processes, IntegralGP covariance functions, volumetric regression, heterogeneous spatial supports, data fusion, ore grade estimation, mine geology modeling innovation. \textbf{MSC}: 68U99 (Computing methodologies and applications: misc.), 60G15 (Gaussian processes), 62G08 (Nonparametric regression)
}\\
\end{tabular}
\thispagestyle{specialfooter}

\newpage
\section{Introduction}\label{sec:intro}
Spatial support refers to a specific volume, area or interval over which data is collected or analysed in spatial statistics. It plays a crucial role in determining the accuracy of estimations, especially when it comes to predicting subterranean properties. The inference grids (or blocks) specified for an inference model are often incompatible with the size of the input samples, furthermore, predictions are generally obtained using some combination of data sources which are measured at different scales with different levels of uncertainties \citep{casttrignano2020spatialsupport}. Data fusion refers to mathematical approaches that fundamentally enable these differences to be reconciled. According to \citet{hall2004mathematical}, it is the process of ``combining information from heterogeneous sources into a single composite picture of the relevant process, such that the composite picture is generally more accurate and complete than that derived from any single source alone.''

\begin{figure*}[h]
\centering
\includegraphics{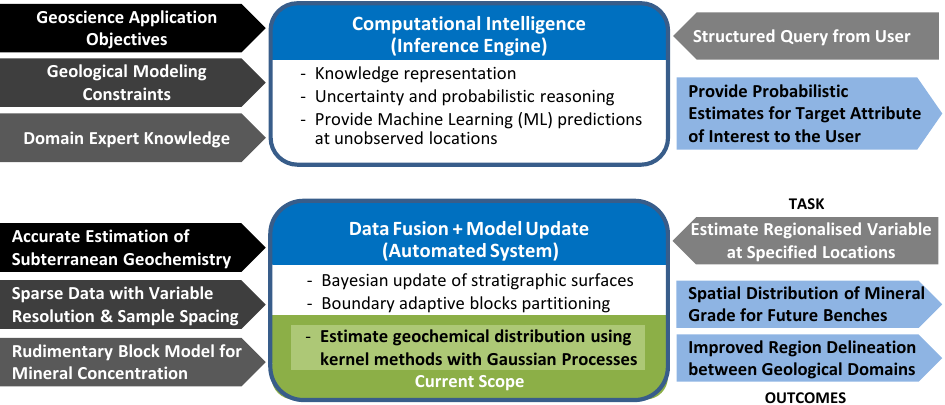}
\caption{Top: General architecture of an expert system commonly encountered in environmental and geo-sciences. Bottom: Specific context for the mining domain expert system described in this paper.}\label{fig:mining-geology-expert-system}
\end{figure*}

This article describes a data fusion algorithm that has applications to mineral resource estimation in mining engineering and geology. At a high level, it follows the general architecture of expert systems commonly encountered in environmental and geological sciences. A central characteristic of such expert system is its intense focus on addressing some geoscientific application objectives, and that it takes into account modelling constraints and domain expert knowledge, as depicted in the top half of Fig.~\ref{fig:mining-geology-expert-system}. Knowledge representation, uncertainty and probabilistic reasoning are the critical elements that endow the system with computational intelligence, and these constitute what data scientists often refer to as the \textit{inference} engine. The most common usage scenario involves receiving a structured query from users as an input, for instance, the system might be tasked to make predictions at unobserved locations using machine learning techniques, then provide probabilistic interpretations for a target attribute of interest to the user as an output.

In the bottom half of Fig.~\ref{fig:mining-geology-expert-system}, these concepts are mapped to concrete ideas to provide an industry perspective and a context for the mining domain expert system described in this work. It begins with the objective of obtaining accurate estimation of the subterranean geochemical composition. Although this knowledge can conceivably be used to implement a cost-efficient adaptive sampling strategy based on reinforcement learning, the envisaged system is constrained to use only fixed measurements at this time. Without loss of generality, the input consists of chemical assay data with variable resolution and sample spacing. In addition, expert knowledge of the geological structure---which informs the mineral composition and spatial distribution of geozones---is encapsulated in a rudimentary model that prescribes the location and attribute value of individual blocks that span the modelled region. The data fusion component described in this article is an extension of \citep{melkumyan2011fusion} and is best viewed as part of a larger system that automatically updates an orebody model \citep{guo2022automatic}. In previous works, the authors have described a Bayesian method for warping/morphing stratigraphic surfaces. The goal was to improve geological boundary definitions and reconcile the model with geochemical evidence obtained from blasthole assays \citep{lowe2021bayesian}. These incremental updates are coupled with a spatial restructuring algorithm \citep{leung2020mos} that re-partitions the blocks to localise and align with the adjusted surfaces. This helps reduce smoothing distortion when attributes are interpolated across geological boundaries or ore grade discontinuities. Coming full circle, this work describes a novel and versatile kernel-based framework for estimating geochemical properties in an ore deposit. The task is to estimate a regionalised variable (e.g., the concentration of iron) at specified locations. Owing to the top-down manner in which surface mining operates in an open-pit mine, two outcomes are especially cherished. First, the ability to faithfully predict the spatial distribution of mineral in the bench below; the caveat is using only historical data and supplementary blasthole data available from the current operating bench. Second, the ability to improve region delineation and separate high-grade ore and low-grade/waste material between geological domains. These outcomes have a positive flow-on effect, they can improve the quality of material tracking and grade control \citep{jupp2013role} associated with ore excavation and portfolio optimisation \citep{navarro2019innovative,chamanbaz2025stoch}. These issues are further discussed along with future challenges and opportunities for mining automation in a survey paper \citep{leung2023survey}.

The strength of the present proposal lies not only in utilising multiple sources of data, but catering for measurements with different spatial supports in a principled way, instead of treating all samples simply as points \citep{chlingaryan2024augment}. As motivation, we elaborate on an application scenario in the next section and show that both outcomes (more accurate regression and boundary delineation) are significantly improved when the spatial support of measurements are properly considered. Thus far, the key impediments for progress have been a lack of awareness of the differences this formulation can make, and the required changes to the kernels (covariance functions) that would allow these benefits to be realised. In this contribution, the relevant mathematical expressions are presented and some of the practical issues such as numerical stability are discussed.

\subsection{Contributions and Organisation}\label{sec:contributions}
To summarise, this work broadly relates to the inference component in a mining-oriented expert system, where computational intelligence is concerned with knowledge representation and probabilistic reasoning. The expert system may be viewed as an intelligent system that provides data fusion and model update capability using machine learning. The objective is to provide improved grade estimation beneath the drilled bench; these in turn can facilitate better grade control and mine planning. The main proposal revolves around the Integral Gaussian Process (IntegralGP) framework which enables volumetric estimation of subterranean geochemical properties in a mineral deposit.

This paper makes two contributions: 1) a volumetric regressor that makes probabilistic grade prediction using IntegralGP; and 2) demonstration of a data fusion algorithm for grade modelling in an open-pit iron ore deposit. The novelty of the IntegralGP formulation is that it takes into consideration the different spatial supports of data, such as volumetric and interval measurements encountered in mine geology modelling. In current practice, these chemically compatible but spatially incongruous samples are typically all treated as point measurements \citep{monjezi2013comparative,jafrasteh2018comparison,leung2024porphyry}. In this work, we show that properly accounting for the spatial dimensions of these measurements can lead to lower regression error and better boundary delineation between low-grade and high-grade ore zones. Viewing through these lens, the first contribution is grounded in theoretical development. IntegralGP describes mathematical changes to the covariance functions that allow these benefits to be realised. The second contribution deals with application, whereby sparse data with variable resolution and sample spacing are combined. Specifically, a supplementary data source is used to improve the grade prediction of an existing long-range block model.

In terms of organisation, Sec.~\ref{sec:background} uses motivating examples to illustrate measurements with different spatial supports. It lays the foundation for understanding the relevant issues and merits of our proposal. Building on the concepts of kernels and GP regression, Sec.~\ref{sec:methods} presents the IntegralGP framework with emphasis on the formulation of the covariance functions. Following theoretical development, Sec.~\ref{sec:application} describes the data fusion algorithm and gradually shifts the focus to applications. Experiment results are presented in Sec.~\ref{sec:experiment-results} with the aims of shedding light in three areas: data assimilation, model validation and material classification. The first discussion illustrates data fusion using blasthole assays and how heteroscedastic GP (noise level) works in the proposed algorithm. The second highlights differences visually and quantifies improvement in regression performance. The third examines the mean absolute error and model bias and treats the predicted grade as a categorical variable. Concluding remarks will follow in Sec.~\ref{sec:conclusion}. Finally, research perspectives and practicalities are considered in Appendices~\ref{app:research-perspective} and \ref{app:practicalities}.

\section{Background}\label{sec:background}
The goal of this section is to lay a strong foundation for understanding the relevant issues and demonstrate the merits of our proposal. Although it is legitimate to say it is technically incorrect to model interval measurements as points, this line of reasoning is so reductive, readers may find it uneasy to accept this at face value. To convince readers of the benefits and motivate this study, we examine an exemplary scenario in Sec.~\ref{sec:motivation} and use it to explain what difference IntegralGP can make. This provides targeted evidence on how the proposed methodology can improve regression and classification outcomes. After establishing the significance of this paradigm shift, we present the kernel based Gaussian Process (GP) framework in Sec.~\ref{sec:gp} to familiarise readers with the computational engine in the expert system before making any alteration. Subsequently, we describe the innovative part of the system and develop the mathematics which enable the objectives---fusion of heterogeneous data, more accurate regression and boundary delineation---to be achieved in Sec.~\ref{sec:methods}.

\subsection{Motivation}\label{sec:motivation}
\begin{figure*}[h]
\centering
\includegraphics{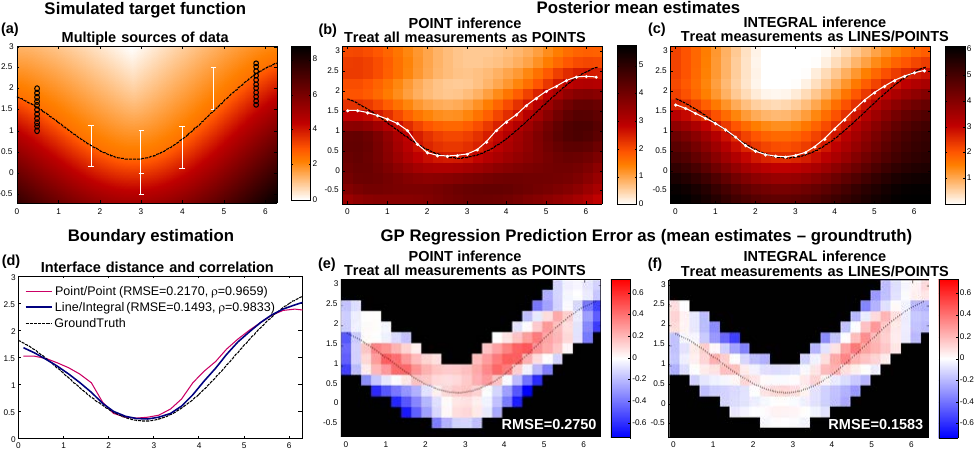}
\caption{A motivating example illustrating the potential benefits of integral GP. Computational details are given in Appendix~\ref{app:construction}.}\label{fig:motivation1}
\end{figure*}
Figure~\ref{fig:motivation1}a illustrates a folding mineralisation scenario which is commonly encountered in major iron ore deposits. It depicts a geochemical gradient whereby the mineral concentration transitions from low to high (graded from waste to ore from the top down) across the stratigraphic boundary (black curve). The general objective is to estimate the simulated target function which in practice is unknown using only sparse assay samples. While short-interval measurements (typically 1--2m in length) can be treated reasonably as points, the point of contention is whether longer-interval measurements can still be treated as points, since blasthole assays are typically taken over an interval of $10$m or more. In panel (a), two sources of data are available. Point measurements have high vertical precision but are spaced large distances apart; these are represented by black open circles. Line measurements represent the average grade over a finite interval. These samples have much lower vertical resolution but are sampled more densely in the horizontal plane; they are represented by white bars. The next two panels compare the regression results for point based GP and integral GP. For the status quo, both measurements and inference locations are treated as points in (b). For the proposed method, line measurements and inference grids are honoured as intervals and blocks, respectively. Hence, the mean grade is estimated in (c) with proper regard for their spatial support dimensions. Extracting the level-set at threshold 2.667, the resultant boundary estimates are shown as white dotted lines in (b) and (c). The first observation from (d) is that the boundary extracted by integral GP is more accurate; the groundtruth distance as measured by RMSE are 0.1493 and 0.2170 for Line/Integral and Point/Point, respectively. The second observation from (e) and (f) is that the values predicted by integral GP have lower distortion (with RMSE of 0.1583 and 0.2750 for Line/Integral and Point/Point respectively) when performance is evaluated near the interface where samples are available.

\begin{figure*}[h]
\centering
\includegraphics{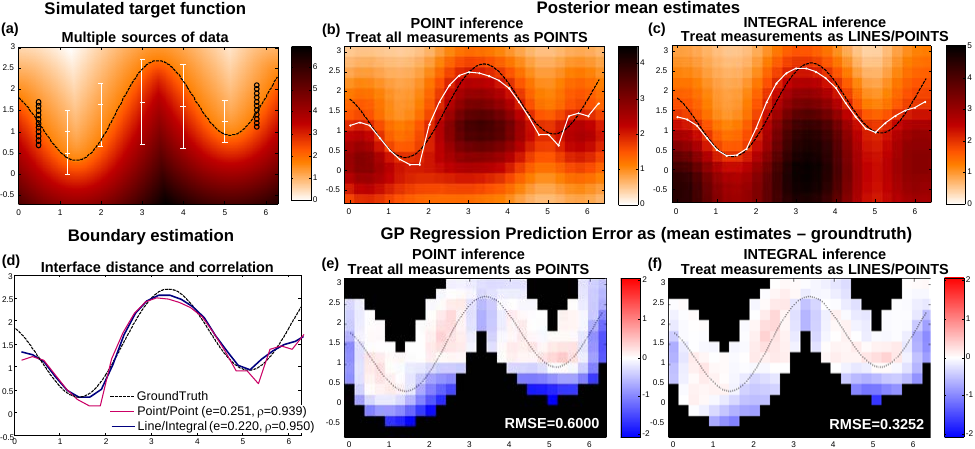}
\caption{A second example illustrating the potential benefits of integral GP.}\label{fig:motivation2}
\end{figure*}

Similar observations also hold in more challenging scenarios, such as the syncline-anticline topology shown in Fig.~\ref{fig:motivation2} where the structure changes at a rate that is twice as fast. In this case, the boundary distortion is greater and the ore/waste region delineation performance is worse under point GP when line measurements are simply treated as points. These examples made clear the reasons for pursuing integral GP.

\subsection{Gaussian process regression: a kernel based framework}\label{sec:gp}
To familiarise with the technology that underpins the inference engine, the Gaussian Process (GP) kernel framework will now be described to provide clarity on how the posterior mean estimates seen in Fig.~\ref{fig:motivation2}b are obtained. The general problem involves predicting a target attribute $y_*\in\mathbb{R}$ at queried locations $\vec{x}_*\in\mathbb{R}^D$ given training data (or measurements) $\mathcal{D}=\{(\vec{x}_i,y_i)\}_{i=1:N}$ where $y_i$ is known. The asterisk indicates the variable belongs to the test set, where $y_*=f(\vec{x}_*)$ for some unknown random function $f$.  In machine learning literature, the input observations are often grouped together, so that the training data may be written in matrix notation as $\mathcal{D}=\{\mat{X},\vec{y}\}$ with $\mat{X}=[\vec{x}_1,\ldots,\vec{x}_N]\in\mathbb{R}^{D\times N}$ and $\vec{y}\in\mathbb{R}^N$.

When regression is performed using GP, a Gaussian prior is placed over the function $f$. Using the observations in $\mathcal{D}$, Bayesian inference infers $p(f\!\mid\! \mathcal{D})$ and converts this into a posterior via the following equation \citep{williams2006gaussian}
\begin{equation}
p(y_*\!\mid\! \mat{X}_*, \{\mat{X},\vec{y}\})=\int p(y_*\!\mid\! f,\mat{X}_*)p(f|\mathcal{D})df\label{eq:bayesian-inference}
\end{equation}
A Gaussian process may be viewed as a collection of random variables, $f(\vec{x})\sim\text{GP}(m(\vec{x}), k(\vec{x},\vec{x}'))$, which are individually and jointly Gaussian distributed. Thus, a GP is completely specified by a mean function $m(\vec{x})=\mathbb{E}[f(\vec{x})]$ and covariance function $k(\vec{x},\vec{x}')=\mathbb{E}[(f(\vec{x})-m(\vec{x}))(f(\vec{x}')-m(\vec{x}'))^T]$. GP regression defines a joint distribution $p(\vec{f}\!\mid\! \mat{X})=\mathcal{N}(\vec{f}\!\mid\! \vec{\mu},\mat{K})$ over a set of $N$ data points, where $\mat{K}_{ij}=k(\vec{x}_i,\vec{x}_j)$ and $\vec{\mu}=[m(\vec{x}_1),\ldots,m(\vec{x}_N)]^T$. Here, $k$ represents a positive semi-definite kernel (covariance function) that models spatial correlation in the data. For a stationary process, $k(\vec{x}_i,\vec{x}_j)=k(\vec{x}_j,\vec{x}_i)=k(\vec{d})$ represents a smooth decaying symmetric function that depends only on the displacement $\vec{d}=\vec{x}_i-\vec{x}_j$. If the kernel is isotropic, then one can simply write $k(d)$ as it depends only on the lag distance $d=\Vert\vec{d}\Vert$. The implicit assumption is that $f(\vec{x}_i)$ and $f(\vec{x}_j)$ will be more similar when $\vec{x}_i$ and $\vec{x}_j$ are closer together.

For noisy observations, the training data is assumed to satisfy $y_i=f(\vec{x}_i)+\epsilon_i$ where $\epsilon_i\sim\mathcal{N}(0,\sigma_y^2)$. The joint density governing the observed and test data is given by
\begin{align}
\begin{bmatrix}\vec{y}\\\vec{f}_*\end{bmatrix}\sim\mathcal{N}\left(0, \begin{bmatrix}\mat{K}_y & \mat{K}_*\\ \mat{K}_*^T & \mat{K}_{**}\end{bmatrix}\right)
\label{eq:gp-joint-density}
\end{align}
and the posterior predictive density is given by
\begin{align}
p(\vec{f}_*\!\mid\! \mat{X}_*,\mat{X},\vec{y})=\mathcal{N}(\vec{f}_*\!\mid\! \vec{\mu}_*,\mat{\Sigma}_*)
\label{eq:gp-posterior-density}
\end{align}
It is worth emphasising that GP regression produces a probabilistic prediction \citep{leung2024porphyry} at each queried location $\vec{x}_*$ in the form of $p(\vec{f}_*\!\mid\! \mat{X}_*,\mat{X},\vec{y})$. Equations that describe the prediction outcomes, viz., the posterior mean, $\vec{\mu}_*$, and covariance, $\mat{\Sigma}_*$, are given by
\begin{align}
\vec{\mu}_* &= \mathbf{\mu}(\mat{X}_*)+\mat{K}_*^T \mat{K}_y^{-1}(\vec{y}-\mathbf{\mu}(\mat{X}))\label{eq:gp-posterior-mean}\\
\mat{\Sigma}_* &= \mat{K}_{**}-\mat{K}_*^T \mat{K}_y^{-1} \mat{K}_*\label{eq:gp-posterior-cov}\\
\mat{K}_y &= \text{cov}(\vec{y}\!\mid\! \mat{X})=\mat{K}+\sigma_y^2 \mat{I}\label{eq:gp-training-data-cov-matrix}
\end{align}
The training data covariance matrix $\mat{K}\in\mathbb{R}^{N\times N}$ is independent of $\vec{x}_*$ whereas $\mat{K}_*\in\mathbb{R}^{N\times M}, \mat{K}_*[i,j]=k(\vec{x}_i, \vec{x}_{*j})$ and $\mat{K}_{**}\in\mathbb{R}^{M\times M}, \mat{K}_{**}[i,j]=k(\vec{x}_{*i}, \vec{x}_{*j})$ both depend on the queried locations, $\vec{x}_*$. The matrix $\mat{K}_*$ satisfies the transposition property $\mat{K}^*=\mat{K}_*^T$ since $\mat{K}^*[i,j]\defeq k(\vec{x}_{*i},\vec{x}_j)=\mat{K}_*[j,i]$. For mineral resource estimation, an efficacious choice for the covariance function is the Mat\'ern 3/2 kernel \citep{melkumyan2011multi} whose functional form is given in Table~\ref{tab:varphi-grad} alongside other options.

GP training refers to the process of fitting kernel hyperparameters, $\vec{\theta}=[\sigma_f^2,l_1,l_2,\ldots,l_D,\sigma_y^2]$, to the training data (assuming $\vec{x}\in\mathbb{R}^D$). When $D=3$, $\sigma_f^2$, $[l_1,l_2,l_3]$ and $\sigma_y^2$ represent the squared amplitude, x, y and z kernel length scales and noise power, respectively. The training objective is to maximise the log marginal likelihood (LML) with respect to $\vec{\theta}$:
\begin{align}
\log p(\vec{y}\!\mid\! \mat{X},\vec{\theta})=-\frac{1}{2}\vec{y}^T\left[\mat{K}+\sigma_y^2 \mat{I}\right]^{-1}\vec{y} - \frac{1}{2}\log\left|\mat{K}+\sigma_y^2 \mat{I}\right|-\frac{N}{2}\log 2\pi\label{eq:gp-lml}
\end{align}
where $\left|\,\cdot\,\right|$ represents the determinant. The marginal likelihood in (\ref{eq:gp-lml}) contains three terms that represent (from left to right) the data fit, complexity penalty and normalisation constant. As the marginal likelihood is a non-convex function of the hyperparameters \citep{lalchand2020approximate}, only local maxima can be obtained. In practice, gradient descent algorithms are usually employed for this optimisation using multiple starting points \citep{melkumyan2009sparse}. In Sec.~\ref{sec:methods}, the novel aspects of this proposal will be described to shed light on the modifications needed to implement integral GP and the data fusion algorithm.

\section{Methods}\label{sec:methods}
This section presents a unified representation that enables heterogeneous data with compatible geochemistry but different spatial supports to be modelled through single-task GP. In particular, point-based locations (see $\mat{X}$ and $\mat{X}_*$ in (\ref{eq:gp-posterior-density})) are extended to include sample dimensions, and the relevant changes to the covariance functions for point, interval and volumetric combinations are described. To assist with practical deployment, formulas for several kernels in the exponential and Mat\'ern families are given and some empirical observations relating to numerical stability are discussed.

\subsection{Heterogeneous Data}\label{sec:hetero-data}
For demonstration, this paper describes the assimilation of two data sources in a mining geology context. The first source, referred as EPR, is a regularised block model created by domain experts using reverse circulation (RC) drilling and core samples collected a priori during the mining exploration phase. RC drilling flushes the drilled material to the surface using compressed air, field samples are gathered at regular intervals and sent to laboratory for geochemical analysis. This type of assay data is sparse (often spaced hundreds of meters apart) but has high vertical resolution ($\le 1$m). The EPR is spatially represented by a uniform array of blocks, typically $15\times 15\times 10$~m\textsuperscript{3} in size, each accompanied by an average grade (e.g. the concentration of iron as a weight percentage) typically estimated via kriging. The EPR model provides a fair estimate of the underlying geochemical distribution at large scales but it is considered as locally inaccurate. The main reason is that geologists have a direct input in ensuring the general geological structure is captured in the model, however there is considerable uncertainty at the local scale due to the sparsity of the measurements available when the model was created. The second source relates to blasthole (BH) samples gathered subsequently during the mining production phase \citep{engstrom2017optimal}. Typically, X-ray fluorescence is used to determine the geochemical signature/material composition of the sample, which is then attributed to a subterranean interval at a certain depth along the drilled hole. Hence, EPR and BH samples may be treated respectively as volumetric data and line measurements of the average grade.

\subsection{Observation model}\label{sec:observ-model}
To provide a unified description of the heterogeneous data, let $A_i=[a_{i,1},a_{i,2},a_{i,3}]^T$ and $H_i=[h_{i,1},h_{i,2},h_{i,3}]^T$ denote the centroid (x, y, z coordinates) and dimensions (length, width, height) of the $i^\text{th}$ sample. This is sufficient for representing point, line and volumetric measurements, noting for instance that the special case $h_{i,1}=h_{i,2}=0$ and $h_{i,3}\neq 0$ would cater for blasthole samples. This perspective allows spatially dissimilar, single-species chemically correlated measurements (e.g. Fe grade values obtained from EPR and BH) to be modelled jointly using a single Gaussian Process (GP) without resorting to multi-task GP \citep{vasudevan2012information}. The problem may be formulated as follows.

Let $f(\vec{x}):\mathbb{R}^D\rightarrow\mathbb{R}$ be a GP regression model that predicts the attribute at $\vec{x}\in\mathbb{R}^D$. Suppose the training data comprises noisy observations $\vec{y}$ at $N$ input locations $\mat{X}$ as shown in (\ref{eq:training-set-coords})--(\ref{eq:observ-noise})
\begin{align}
\mat{X}&=\begin{bmatrix}A_1 & A_2 & \ldots & A_N\\H_1 & H_2 & \ldots & H_N\end{bmatrix}\in\mathbb{R}^{2D\times N}\label{eq:training-set-coords}\\
\vec{y}&=[y_1, y_2, \ldots, y_N]\label{eq:training-set-values}\\
y_i&=u_i+\varepsilon_i,\quad u_i=\dfrac{1}{\left| V_i\right|}\int_{V_i}f(\vec{x}) d\vec{x}\label{eq:observ-noise}
\end{align}
where $u_i$ denotes the noise-free component of a volumetric measurement, $\varepsilon_i\sim N(0,\sigma_i^2)$ denotes Gaussian noise with zero mean and variance $\sigma_i^2$. The inference task is to predict the value at each queried location $X_{*,j}=\begin{bmatrix}A_{*,j}\\H_{*,j}\end{bmatrix}$ by computing $y_{*,j}=\dfrac{1}{\left| V_{*,j}\right|}\int_{V_{*,j}}f(\vec{x}_*) d\vec{x}_*$. This integral is averaged over the volume of block $j$ in the EPR model, or interval $j$ of a blasthole. Without loss of generality, $V_i$ in (\ref{eq:observ-noise}) can represent volumes, planes, lines and points; these special cases are further considered in Sec.~\ref{sec:cov-vols-lines-pts}. Throughout this paper, the asterisk notation serves as a hint that we are referring to an inference location where the true attribute value is unknown.

\subsection{Covariance between function and its integral}\label{sec:cov-fn-integral}
Writing $\vec{s}=\vec{x}_*$ to simplify the notation, the objective is to show the covariance between $f(\vec{x})$ at any location $\vec{x}$ and $\int_\Omega f(\vec{s}) d\vec{s}$ at a different location $\vec{s}$ over some region of integration $\Omega$ can be expressed in the form of (\ref{eq:cov-integral-fs-fx}).
\begin{align}
\textbf{Lemma:}\quad\cov{\left(\int_\Omega f(\vec{s})d\vec{s}, f(\vec{x})\right)}=\int_\Omega k(\vec{s},\vec{x})d\vec{s}\quad\text{where }k(\vec{s},\vec{x})=\cov(f(\vec{s}),f(\vec{x}))\label{eq:cov-integral-fs-fx} 
\end{align}

This result follows directly from Fubini's theorem \citep{veraar2012stochastic}.\footnote{The key observation is that $\cov{\left(\int_\Omega f(\vec{s})d\vec{s}, f(\vec{x})\right)}\defeq\mathbb{E}\left[\left(\int_\Omega f(\vec{s})d\vec{s}-\mathbb{E}\left[\int_\Omega f(\vec{s})d\vec{s}\right]\right) \left(f(\vec{x})-\mathbb{E}\left[f(\vec{x})\right]\right)\right]$ equals $\int_\Omega\mathbb{E}\left[\left(f(\vec{s})-\mathbb{E}\left[f(\vec{s})\right]\right) \left(f(\vec{x})-\mathbb{E}\left[f(\vec{x})\right]\right)\right] d\vec{s}$.} Its significance is that the spatial relationship between measurements with different spatial supports is basically given by an integral of their covariance function $\int_\Omega k(\vec{x}_*,\vec{x})$, where $k(\cdot,\cdot)$ represents a generic kernel. The detailed expressions will be presented in the next section, accounting for differences in their spatial supports. This would enable, for instance, the correlation between a blasthole sample (line measurement) at $\vec{x}_i$ and a query location (volumetric block) $\vec{x}_{*,j}$ in the EPR model to be computed.

\subsection{Covariance between volumes, lines and points}\label{sec:cov-vols-lines-pts}
To obtain covariance functions for measurements with different spatial supports, we will assume the kernel is separable (can be factorised in product form as in (\ref{eq:separable-kernel})) and derives from a once differentiable function $\Phi$ and twice differentiable function $\Psi$ in (\ref{eq:kernel-differentiability}). In (\ref{eq:separable-kernel}), $l_m$ denotes the length scale hyperparameter along the corresponding axis.
\begin{equation}
\cov\left(f(\vec{x}),f(\vec{s})\right)=k(\vec{x},\vec{s})=\prod_{m=1}^{D}\varphi\left(\dfrac{x_m-s_{m}}{l_m}\right)\label{eq:separable-kernel}
\end{equation}
\begin{align}
\varphi(t)&=\dfrac{d\Phi}{dt}=\Phi'\label{eq:kernel-differentiability}\\
&=\dfrac{d^2\Psi}{dt^2}=\Psi''\notag
\end{align}

The general covariance expressions shown in (\ref{eq:cov-yi-yj})--(\ref{eq:rho}) are obtained by the fundamental theorem of calculus from (\ref{eq:cov-integral-fs-fx}), (\ref{eq:separable-kernel}) and (\ref{eq:kernel-differentiability}). The appearance of the Dirac delta function [$\delta_{ij}=1$ if $i=j$, 0 otherwise] in (\ref{eq:cov-yi-yj}) is the consequence of assuming the signal and noise components in (\ref{eq:observ-noise}) to be uncorrelated between two separate measurements.

\begin{align}
\cov(y_i,y_j) &= \cov(u_i,u_j)+\sigma_i^2\delta_{ij}\label{eq:cov-yi-yj}\\
\cov(u_i,u_j) &= \cov(y_i,u_j) = \cov(u_i,y_j)\notag\\
&= \prod_{m=1}^{D}\begin{cases}
\dfrac{l_m^2}{h_{i,m}h_{j,m}}R(a_{i,m},a_{j,m},h_{i,m},h_{j,m},l_m), & \text{if }h_{i,m}\neq 0, h_{j,m}\neq 0\\
\dfrac{l_m}{h_{j,m}}\rho(a_{j,m},h_{j,m},a_{i,m},l_m), & \text{if }h_{i,m}=0, h_{j,m}\neq 0\\
\dfrac{l_m}{h_{i,m}}\rho(a_{i,m},h_{i,m},a_{j,m},l_m), & \text{if }h_{i,m}\neq 0, h_{j,m}=0\\
\varphi\left(\dfrac{a_{i,m}-a_{j,m}}{l_m}\right), & \text{if }h_{i,m}=0, h_{j,m}=0\\
\end{cases}\label{eq:cov-yi-uj}
\end{align}
where
\begin{align}
R(a_1,a_2,h_1,h_2,l)=&-\Psi\left(\frac{a_1-a_2+(h_1-h_2)/2}{l}\right)+\Psi\left(\frac{a_1-a_2-(h_1+h_2)/2}{l}\right)\notag\\
&+\Psi\left(\frac{a_1-a_2+(h_1+h_2)/2}{l}\right)-\Psi\left(\frac{a_1-a_2-(h_1-h_2)/2}{l}\right)\label{eq:R}\\
\rho(a,h,x,l)=&\,\Phi\left(\frac{a+\frac{h}{2}-x}{l}\right)-\Phi\left(\frac{a-\frac{h}{2}-x}{l}\right)\label{eq:rho}
\end{align}
Note: $l$ stands for vertical line and is used for BH assays.

\noindent For instance, line-volume covariance would result in expression (\ref{eq:cov-line-volume}) from (\ref{eq:cov-yi-uj}).
\begin{align}
\resizebox{0.93\textwidth}{!}{$\cov(X_L,X_V)=\dfrac{l_1}{h_{V,1}}\rho(a_{V,1},h_{V,1},a_{L,1},l_1) \dfrac{l_2}{h_{V,2}}\rho(a_{V,2},h_{V,2},a_{L,2},l_2) \dfrac{l_3^2}{h_{L,3}h_{V,3}}R(a_{L,3},a_{V,3},h_{L,3},h_{V,3},l_3)$}\label{eq:cov-line-volume}
\end{align}

\subsection{Partial derivatives for learning hyperparameters}\label{sec:cov-partial-derivatives}
Since the partial derivatives of the covariance function are used to maximise the log marginal likelihood (\ref{eq:d-lml-d-theta}) during GP learning \citep{williams2006gaussian} to optimise the hyperparameters $\vec{\theta}=(\theta_1,\ldots,\theta_D)$, the spatial support dependent partial derivatives must be obtained. These are considered for three cases, between (a) a pair of volumes, (b) a volume and point or vice-versa, and (c) a pair of points.
\begin{align}
\dfrac{\partial\log p(\vec{y}\!\mid\!\mat{X},\vec{\theta})}{\partial\theta_q}&=\dfrac{1}{2}\vec{y}^T \mat{K}^{-1} \dfrac{\partial\mat{K}}{\partial\theta_q}\mat{K}^{-1}\vec{y} - \dfrac{1}{2}\text{tr}\left(\mat{K}^{-1}\dfrac{\partial\mat{K}}{\partial\theta_q}\right)\label{eq:d-lml-d-theta}\\
&=\dfrac{1}{2}\text{tr}\left((\vec{\alpha}\vec{\alpha}^T-\mat{K}^{-1})\dfrac{\partial\mat{K}}{\partial\theta_q}\right)\text{ where }\vec{\alpha}=\mat{K}^{-1}\vec{y}, \mat{K}=[k(\vec{x}_i,\vec{x}_j)]_{i,j}\notag
\end{align}

\noindent When $h_{i,q}\neq 0$ and $h_{j,q}\neq 0$,
\begin{align}
&\dfrac{\partial\cov(u_i,u_j)}{\partial l_q}=\cov(u_i,u_j)\left[\dfrac{2}{l_q}+\dfrac{1}{R(\cdot)}\left(\lambda_{(h_{i,q}-h_{j,q})}^{\Phi} - \lambda_{-(h_{i,q}+h_{j,q})}^{\Phi} - \lambda_{(h_{i,q}+h_{j,q})}^{\Phi} + \lambda_{-(h_{i,q}-h_{j,q})}^{\Phi}\right)\right]\label{eq:partial_hiq1_hjq1}\\
&\text{where }\lambda_{h}^{\Phi}=\dfrac{a_{i,q}-a_{j,q}+\frac{h}{2}}{l_q^2}\Phi\left(\dfrac{a_{i,q}-a_{j,q}+\frac{h}{2}}{l_q}\right)
\end{align}
\noindent When $h_{i,q}=0$ and $h_{j,q}\neq 0$,
\begin{align}
&\dfrac{\partial\cov(u_i,u_j)}{\partial l_q}=\cov(u_i,u_j)\left[\dfrac{1}{l_q}+\dfrac{1}{\rho(\cdot)}\left(\lambda_{a_{j,q}-a_{i,q},\,-h_{j,q}}^{\varphi} - \lambda_{a_{j,q}-a_{i,q},\,h_{j,q}}^{\varphi}\right)\right]\label{eq:partial_hiq0_hjq1}\\
& \text{where }\lambda_{a,h}^{\varphi}=\dfrac{a+\frac{h}{2}}{l_q^2}\varphi\left(\dfrac{a+\frac{h}{2}}{l_q}\right)
\end{align}
\noindent When $h_{i,q}\neq 0$ and $h_{j,q}=0$,
\begin{align}
&\dfrac{\partial\cov(u_i,u_j)}{\partial l_q}=\cov(u_i,u_j)\left[\dfrac{1}{l_q}+\dfrac{1}{\rho(\cdot)}\left(\lambda_{a_{i,q}-a_{j,q},\,-h_{i,q}}^{\varphi} - \lambda_{a_{i,q}-a_{j,q},\,h_{i,q}}^{\varphi}\right)\right]\label{eq:partial_hiq1_hjq0}
\end{align}
\noindent When $h_{i,q}=0$ and $h_{j,q}=0$,
\begin{align}
\dfrac{\partial\cov(u_i,u_j)}{\partial l_q}&=-\cov(u_i,u_j)\left[\dfrac{a_{i,q}-a_{j,q}}{l_q^2\,\varphi\left(\dfrac{a_{i,q}-a_{j,q}}{l_q}\right)}\varphi'\left(\dfrac{a_{i,q}-a_{j,q}}{l_q}\right)\right]\label{eq:partial_hiq0_hjq0}
\end{align}

\subsection{Expressions for kernels, gradients and anti-derivatives}\label{sec:kernels}
The positive semi-definite function $\varphi$ in (\ref{eq:separable-kernel}) corresponds to a GP kernel. For several commonly used kernels, the formula for $\varphi$, its gradient $\varphi'$ and anti-derivatives $\Phi$ and $\Psi$ as cited in (\ref{eq:R}), (\ref{eq:rho}), (\ref{eq:partial_hiq1_hjq1})--(\ref{eq:partial_hiq0_hjq0}) are given in Tables~\ref{tab:varphi-grad} and \ref{tab:Phi-Psi}.

\begin{table*}[t]
\centering
\def\arraystretch{1.35}
\resizebox{0.9\textwidth}{!}{
\begin{tabular}{lcc}\hline
{\small Kernel family} & $\varphi(t)$ & $\varphi'(t)$\\\hline
\\\\[-5\medskipamount] 
{\small Squared exponential} & $e^{-\frac{t^2}{2}}$ & $-t e^{-\frac{t^2}{2}}$\\
{\small Exponential} & $e^{-\left|t\right|}$ & $-\sgn(t)\,e^{-\left|t\right|}$\\
{\small Mat\'ern $(\nu=3/2)$} & $\left(1+\sqrt{3}\left|t\right|\right)e^{-\sqrt{3}\left|t\right|}$ & $-3t\,e^{-\sqrt{3}\left|t\right|}$\\
{\small Mat\'ern $(\nu=5/2)$} & $\left(1+\sqrt{5}\left|t\right|+\frac{5}{3}t^2\right)e^{-\sqrt{5}\left|t\right|}$ & $\frac{1}{3}e^{-\sqrt{5}\left|t\right|}\left(10t-\sgn(t)\left(5\sqrt{5}t^2+15\left|t\right|\right)\right)$\\\hline
\end{tabular}
}
\caption{$\varphi$ and $\varphi'$ for commonly used kernels}\label{tab:varphi-grad}
\end{table*}

\begin{table*}[t]
\centering
\def\arraystretch{1.35}
\resizebox{0.95\textwidth}{!}{
\begin{tabular}{lcc}\hline
{\small Kernel family} & $\Phi(t)$ & $\Psi(t)$\\\hline
\\\\[-5\medskipamount] 
{\small Squared exponential} & $\sqrt{\frac{\pi}{2}}\erf\left(\frac{t}{\sqrt{2}}\right)$ & $\sqrt{\frac{\pi}{2}}t\cdot\erf\left(\frac{t}{\sqrt{2}}\right)+e^{-\frac{t^2}{2}}$\\
{\small Exponential} & $\sgn(t)\left(1-e^{-\left|t\right|}\right)$ & $\left|t\right|+e^{-\left|t\right|}$\\
{\small Mat\'ern $(\nu=3/2)$} & $\frac{2}{\sqrt{3}}\sgn(t)\left(1-\left(1+\frac{\sqrt{3}}{2}\left|t\right|\right)e^{-\sqrt{3}\left|t\right|}\right)$ & $\frac{2}{\sqrt{3}}\left|t\right|+\left(1+\frac{1}{\sqrt{3}}\left|t\right|\right)e^{-\sqrt{3}\left|t\right|}$\\
{\small Mat\'ern $(\nu=5/2)$} & $\frac{\sgn(t)}{3}\left(\frac{8}{\sqrt{5}}-\left(\frac{8}{\sqrt{5}}+5\left|t\right|+\sqrt{5}t^2\right)e^{-\sqrt{5}\left|t\right|}\right)$ & $\frac{1}{3}\left(\frac{8}{\sqrt{5}}\left|t\right|+\left(3+\frac{7}{\sqrt{5}}\left|t\right|+t^2\right)e^{-\sqrt{5}\left|t\right|}\right)$\\\hline
\multicolumn{3}{c}{\footnotesize Definitions: $\erf(x)=\frac{2}{\sqrt{\pi}}\int_0^{x}e^{-t^2}dt$, $\sgn(x)=\begin{cases}+1 & x>0\\0 & x=0\\-1 & x<0\end{cases}$}
\end{tabular}
}
\caption{$\Phi$ and $\Psi$ for commonly used kernels}\label{tab:Phi-Psi}
\end{table*}

\subsection{Numerical issues}\label{numerical-issues}
The partial derivatives $\partial\cov / \partial l_q$ can misbehave when the values in the $\varphi$, $\Phi$ and $\Psi$ function arguments become very large. To see this, observe that $R$ and $\rho$ both tend to zero in the limit as $\vert a_1-a_2\vert\rightarrow\infty$ in expressions (\ref{eq:R}) and (\ref{eq:rho}). This in turn causes $\cov(u_i,u_j)$ to vanish in (\ref{eq:cov-yi-uj}) and makes $\partial\cov(u_i,u_j) / \partial l_q$ indeterminate in (\ref{eq:partial_hiq1_hjq1})--(\ref{eq:partial_hiq1_hjq0}) as the fraction becomes $0/0$. Therefore, the case of distant volumes needs to be considered separately when two measurements are a large distance apart.

To resolve the issue via algebraic manipulation, we note that the present forms of the $\rho$ and $R$ functions (with dependencies on $\Phi(t)$ and $\Psi(t)$ as reported in Table~\ref{tab:Phi-Psi}) are considered valid when $t$ is contained in the following interval
\begin{equation}
t\in\left[\frac{a_{i,q}-a_{j,q}-\frac{(h_{i,q}+h_{j,q})}{2}}{l_q}, \frac{a_{i,q}-a_{j,q}+\frac{(h_{i,q}+h_{j,q})}{2}}{l_q}\right]\label{eq:t-valid-interval}
\end{equation}
The case of large distances arises when
\begin{equation}
\vert a_{i,q}-a_{j,q}\vert \ge \frac{h_{i,q}+h_{j,q}}{2}\label{eq:t-instability}
\end{equation}
Under this condition, stability can be restored by omitting the leading term that contains $\sgn(t)$ or $\vert t\vert$ from $\Phi(t)$ and $\Psi(t)$, respectively (see Table~\ref{tab:Phi-Psi}). To justify this, it is instructive to examine the expression $\Phi(t)=\int\varphi(t) dt$ for the Mat\'ern 3/2 kernel, where
\begin{align}
\Phi(t)&=\frac{2}{\sqrt{3}}\sgn(t)\left(1-\left(1+\frac{\sqrt{3}}{2}\left|t\right|\right)e^{-\sqrt{3}\left|t\right|}\right)+C\notag\\
&=-\frac{2}{\sqrt{3}}\sgn(t)\left(1+\frac{\sqrt{3}}{2}\left|t\right|\right)e^{-\sqrt{3}\left|t\right|}+C_0;\qquad C_0=C+\frac{2}{\sqrt{3}}\sgn(t)\label{eq:Phi-constant}
\end{align}

When $\vert a_{i,q}-a_{j,q}\vert$ dominates under condition (\ref{eq:t-instability}), it solely determines the sign of the signum function in (\ref{eq:signum-a-dorminance})
\begin{equation}
\sgn(t)=\sgn\left(\frac{a_{i,q}-a_{j,q}\mp(h_{i,q}-h_{j,q})/2}{l_q}\right)\approx\sgn(a_{i,q}-a_{j,q})\label{eq:signum-a-dorminance}
\end{equation}
Hence, setting $C=-\frac{2}{\sqrt{3}}\sgn(a_{i,q}-a_{j,q})$ results in $C_0=0$. This shows the leading term $\frac{2}{\sqrt{3}}\vert t\vert$ in $\Psi$ (which corresponds to the derivative of the leading term $\frac{2}{\sqrt{3}}\sgn(t)$ in $\Phi$) may be omitted in the case of distant volumes.

\subsection{Partial derivatives for distant volumes}\label{sec:partial-distant-volumes}
As $\varphi(t)$, $\Phi(t)$ and $\Psi(t)$ can be multiplied by the constant $\exp(\vert a_1-a_2\vert/l)$ without changing $\partial\cov(u_i,u_j)/\partial l_q$, an offset may be added to the exponent to stabilise the functions in the case of distant volumes. This is true since the components inside the squared parenthesis in (\ref{eq:partial_hiq1_hjq1})--(\ref{eq:partial_hiq1_hjq0}) remain invariant provided the scaling factors cancel between $\rho(\Phi)$ and $\varphi$, likewise for $R(\Psi)$ and $\Phi$. Under the large distance condition (\ref{eq:t-instability}), the modified expressions for the kernels and anti-derivatives are shown in Table~\ref{tab:modified-varphi-Phi-Psi}.

\begin{table*}[t]
\centering
\def\arraystretch{1.35}
\resizebox{0.95\textwidth}{!}{
\begin{tabular}{lccc}\hline
{\small Kernel family} & $\varphi(t)$ & $\Phi(t)$ & $\Psi(t)$\\\hline
\\\\[-5\medskipamount] 
{\small Exponential} & $e^{-g(t)}$ & $-\sgn(t) e^{-g(t)}$ & $e^{-g(t)}$\\
{\small Mat\'ern 3/2} & $\left(1+\sqrt{3}\left|t\right|\right)e^{-\sqrt{3}g(t)}$ & $-\frac{2}{\sqrt{3}}\sgn(t)\left(1+\frac{\sqrt{3}}{2}\left|t\right|\right)e^{-\sqrt{3}g(t)}$ & $\left(1+\frac{1}{\sqrt{3}}\left|t\right|\right)e^{-\sqrt{3}g(t)}$\\
{\small Mat\'ern 5/2} & $\left(1+\sqrt{5}\left|t\right|+\frac{5}{3}t^2\right)e^{-\sqrt{5}g(t)}$ & $-\frac{\sgn(t)}{3}\left(\frac{8}{\sqrt{5}}+5\left|t\right|+\sqrt{5}t^2\right)e^{-\sqrt{5}g(t)}$ & $\frac{1}{3}\left(3+\frac{7}{\sqrt{5}}\left|t\right|+t^2\right)e^{-\sqrt{5}g(t)}$\\\hline
\multicolumn{4}{c}{\textbf{Substitution:} $g(t)=\vert t\vert - \dfrac{\vert a_1-a_2\vert}{l}$}
\end{tabular}
}
\caption{$\varphi$, $\Phi$ and $\Psi$: modified forms under large distance condition as defined in (\ref{eq:t-instability})}\label{tab:modified-varphi-Phi-Psi}
\end{table*}

\subsection{Summary}\label{sec:methods-summary}
The new kernel definitions that take into account the spatial support of the data do not change the fundamentals in terms of how GP regression operates. The posterior predictive density is still governed by $p(\vec{f}_*\!\mid\!\mat{X}_*,\mat{X},\vec{y})=\mathcal{N}(\vec{f}_*\!\mid\!\vec{\mu}_*,\vec{\Sigma}_*)$ as described in (\ref{eq:gp-posterior-mean})--(\ref{eq:gp-posterior-cov}). What is different is how the elements in the covariance matrices, $\mat{K}[i,j]=k(\vec{x}_i,\vec{x}_j)$, $\mat{K}_*[i,j]=k(\vec{x}_i,\vec{x}_{*j})$ and $\mat{K}_{**}[i,j]=k(\vec{x}_{*i},\vec{x}_{*j})$, are computed. In each case, $k(\vec{x}_i,\vec{x}'_{j})\equiv \cov({u_i,u'_j})$ resolves to a specific product-form expression that depends on the spatial support, $(h_{i,1},h_{i,2},h_{i,3})$ and $(h'_{j,1},h'_{j,2},h'_{j,3})$, according to equation (\ref{eq:cov-yi-uj}). For instance, if $\vec{x}_i$ and $\vec{x}'_j$ represent linear and volumetric measurements, respectively, then it takes on the explicit form given in (\ref{eq:cov-line-volume}). In general, $\cov({u_i,u'_j})$ is a function of the spatial coordinates and sample dimensions, $\vec{a}_i,\vec{a}'_j,\vec{h}_i,\vec{h}'_j\in\mathbb{R}^D$, and it depends on the chosen kernel and its anti-derivatives via $\varphi$, $\rho(\Phi)$ and $R(\Psi)$ which are described in (\ref{eq:R})--(\ref{eq:rho}) and Tables~\ref{tab:varphi-grad} and \ref{tab:Phi-Psi}. The GP kernel hyperparameters are learned by maximising the marginal likelihood (\ref{eq:gp-lml}) using gradient descent techniques. This is accomplished using the partial derivatives given in (\ref{eq:partial_hiq1_hjq1})--(\ref{eq:partial_hiq0_hjq0}). These closed form solutions are expressed in terms of $\lambda^{\Phi}_h$ and $\lambda^{\varphi}_{a,h}$ which once again depend on $\Phi$ and $\varphi$. For samples which are a large distance apart, viz. $\vert a_{i,q}-a'_{j,q}\vert \ge (h_{i,q}+h'_{j,q})/2$, modified expressions are used for numerical stability (see Table~\ref{tab:modified-varphi-Phi-Psi}).

\section{Application}\label{sec:application}
\begin{figure*}[!h]
\centering
\includegraphics{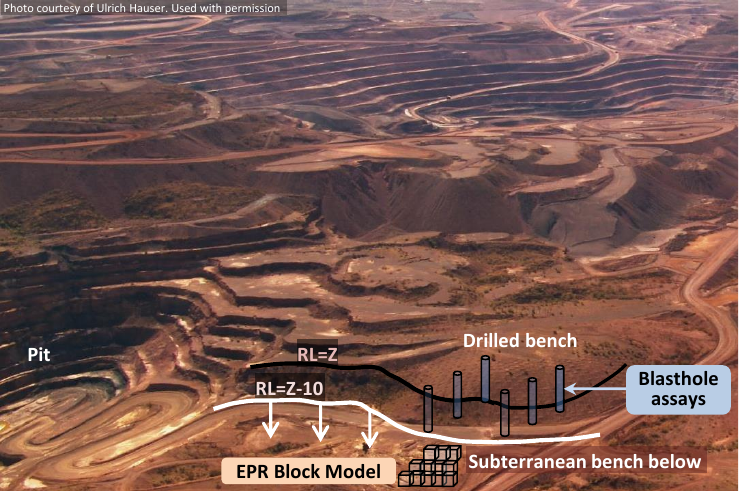}
\caption{Orebody grade modelling as a direct application of the integral GP method. The concept of bench-below prediction and the approximate spatial relationship between blastholes and the EPR block data are depicted in this aerial image of an open-pit iron ore mine. RL denotes reduced level, a surveying terminology that basically means the vertical elevation is expressed with reference to a common frame.}\label{fig:pit-benches}
\end{figure*}

Figure~\ref{fig:pit-benches} illustrates a practical application of the expert system that is consistent with the usage scenario depicted in Fig.~\ref{fig:mining-geology-expert-system}. As mentioned in Sec.~\ref{sec:hetero-data}, the idea is to use the blasthole assay measurements to rectify local inaccuracies in an existing regularised block model (EPR) to obtain improved estimates of the iron ore grade. This requires assimilation of two information sources, viz., the blasthole (BH) line measurements with EPR volumetric data, which will be demonstrated using the proposed integral GP method within a kernel-based regression framework as described in Secs.~\ref{sec:gp}--\ref{sec:partial-distant-volumes}.

\subsection{Fusion algorithm}\label{sec:fusion-algo}
The fusion algorithm is built upon heteroscedastic GP regression which essentially means the noise is modelled as data source or input dependent. In \citet{goldberg1997regression}, Markov Chain Monte Carlo techniques (MCMC) were used to estimate the posterior noise variance, whereas in \citet{kersting2007most}, a faster maximum likelihood approach was proposed and the noise variance was computed using an expectation-maximisation (EM) like iterative optimisation procedure. For this work, the approach presented in \textit{Algorithm}~\ref{alg:fusion} is conceptually similar to \citet{vasudevan2010heteroscedastic} which uses constrained optimisation to determine the noise parameters. Specifically, there are two underlying ideas. First, the target attribute (a regionalised variable pertaining to one chemical species) can be modelled using a single set of GP hyperparameters with just the noise varying between data sets. Therefore, the EPR block model data and blasthole assay samples are viewed as random realisations of a common phenomenon subject to different noise behaviour. Second, the fusion problem is formulated using standard GP precisely as described in Sec.~\ref{sec:gp}, except different data sources have different noise parameters. Thus, the $\sigma^2_y\mat{I}$ term in $\mat{K}_y\!=\!\mat{K}+\sigma^2_y\mat{I}$ in (\ref{eq:gp-training-data-cov-matrix}) is replaced with $\diag\left(\left[n^\mathcal{B}_{1},\ldots,n^\mathcal{B}_{N_B}, n^\mathcal{E}_{1},\ldots,n^\mathcal{E}_{N_E}\right]\right)^2$, where $n^\mathcal{B}_{i}$ and $n^\mathcal{E}_{j}$ denote the specified or estimated noise amplitude for samples $\vec{x}_i$ and $\vec{x}_j$ in the blasthole and EPR data, respectively.

In \textit{Algorithm}~\ref{alg:fusion}, steps 1 and 2 simply identify and represent the data. As indicated in Fig.~\ref{fig:pit-benches}, the blasthole data $\mathcal{D}^\mathcal{B}$ is sourced from the drilled bench (which extends from $Z-10$ to $Z$ meter in elevation) and the EPR regularised block data $\mathcal{D}^\mathcal{E}$ refers to locations in the bench below (from $Z-20$ to $Z-10$ in elevation). The real computation begins in step 3, where GP hyperparameters are learned from blasthole data $\mathcal{D}^\mathcal{B}$ using the covariance and partial derivatives described in Sec.~\ref{sec:methods} which cater for measurements with various spatial supports. In step 4, GP inference is applied to locations in $\mathcal{D}^\mathcal{E}$ to obtain uncertainty estimates $\hat{\sigma}^\mathcal{E}(\vec{x}_{*k})$ for EPR blocks. Since the EPR regularised block model is manually created using data gathered during the mining exploration phase, the noise level associated with each EPR block ($n^\mathcal{E}_k$) is not specified. In step 5, the noise level $n^\mathcal{E}_k$ is inferred from $\hat{\sigma}^\mathcal{E}(\vec{x}_{*k})$ based on the following intuition. When the blasthole uncertainty $\hat{\sigma}^\mathcal{E}(\vec{x}_{*k}$) is low, weighting the evidence toward blasthole observations will yield more confident and reliable predictions. In this situation, the first case in formula (5c) amplifies the noise associated with the EPR value ($n^\mathcal{E}_k$) to diminish its contribution. Conversely, when the blasthole uncertainty $\hat{\sigma}^\mathcal{E}(\vec{x}_{*k})$ is high, the noise associated with EPR ($n^\mathcal{E}_k$) is suppressed---virtually regarded as noise-free---to increase its contribution. In step 6, heteroscedastic GP is applied to the EPR-blasthole combined dataset, the posterior estimates, $\vec{\mu}_*$ and $\vec{\Sigma}_*$, are obtained using the established hyperparameters and EPR noise parameters. In step 6c, the inference blocks $\mat{X}_*$ may represent a denser version of $\mat{X}^\mathcal{E}$ and contain many more cells if the EPR blocks have been further subdivided. This subdivision can be done uniformly or adaptively, for instance, using the spatial restructuring strategy in \citep{leung2020mos} with the goal of localising stratigraphic boundaries, or obtaining a better continuous approximation of the subterranean geochemistry. Ultimately, the key point is that the fusion algorithm provides a trade-off that favours the blasthole source when its information is deemed more reliable by the GP.

\begin{algorithm*}
\begin{algorithmic}[1]
\setcounter{ALG@line}{-1}
\State {\textbf{Input:} $\mathcal{E}$ = EPR regularised block model, $\mathcal{B}$ = blasthole assay samples,}
\NoNumber {\qquad\qquad $Z$ = drilled bench elevation (RL), $\mathcal{K}$ = kernel type (e.g. Mat\'ern 3/2)}
\State Select EPR data $\mathcal{D}^\mathcal{E}$ from the bench of interest and blasthole data $\mathcal{D}^\mathcal{B}$ from the drilled bench above such that
\NoNumber {\quad$\mathcal{D}^\mathcal{E}=\{(A^\mathcal{E}_i,H^\mathcal{E}_i,y^\mathcal{E}_i)\mid a^\mathcal{E}_{i,3}\in[Z-20,Z-10]\}$}
\NoNumber {\quad$\mathcal{D}^\mathcal{B}=\{(A^\mathcal{B}_i,H^\mathcal{B}_i,y^\mathcal{B}_i)\mid a^\mathcal{B}_{i,3}\in[Z-10,Z]\}$. Let $\vert\mathcal{D}^\mathcal{B}\vert=n_B$, $\vert\mathcal{D}^\mathcal{E}\vert=N_E$.}
\State Combine both into a single dataset using the unified representation in (\ref{eq:training-set-coords}) such that
\NoNumber {\quad$\mat{X}=\left[\mat{X}^\mathcal{B}\!\mid\!\mat{X}^\mathcal{E}\right]\in\mathbb{R}^{2D\times(N_B+N_E)}$, $\mat{X}^\mathcal{B}=\begin{bmatrix}A^\mathcal{B}_1 & \ldots & A^\mathcal{B}_{N_B}\\H^\mathcal{B}_1 & \ldots & H^\mathcal{B}_{N_B}\end{bmatrix}$, $\mat{X}^\mathcal{E}=\begin{bmatrix}A^\mathcal{E}_1 & \ldots & A^\mathcal{E}_{N_E}\\H^\mathcal{E}_1 & \ldots & H^\mathcal{E}_{N_E}\end{bmatrix}$.}
\NoNumber {\quad$\vec{y}=\left[\vec{y}^\mathcal{B}\!\mid\!\vec{y}^\mathcal{E}\right]$, where $\vec{y}^\mathcal{B}=[y^\mathcal{B}_1,\ldots,y^\mathcal{B}_{N_B}]$, $\vec{y}^\mathcal{E}=[y^\mathcal{E}_1,\ldots,y^\mathcal{E}_{N_E}]$.}
\State Learn hyperparameters $\vec{\theta}$ by applying GP learning (Sec.~\ref{sec:gp}) to $\mathcal{D}^\mathcal{B}$ with $\mat{X}=\mat{X}^\mathcal{B}$ and $\vec{y}=\vec{y}^\mathcal{B}$. Use the derivatives (\ref{eq:partial_hiq1_hjq1})--(\ref{eq:partial_hiq0_hjq0}) for the covariance function (\ref{eq:cov-yi-uj}) to speed up the optimisation process. Refer to kernel-specific expressions for $\mathcal{K}$ in Tables~\ref{tab:varphi-grad}--\ref{tab:modified-varphi-Phi-Psi}.
\State Using the blasthole training data $\mathcal{D}^\mathcal{B}$, infer the chemical grade (compute $\vec{\mu}_*$ and $\vec{\Sigma}_*$ using (\ref{eq:gp-posterior-mean})--(\ref{eq:gp-posterior-cov})) for blocks $\vec{x}_{*k}\equiv(A^\mathcal{E}_k,H^\mathcal{E}_k,\cdot)\in\mathcal{D}^\mathcal{E}$ in the EPR model. Estimate the blasthole uncertainty $\hat{\sigma}^\mathcal{E}(\vec{x}_{*k})=\sqrt{\vec{\Sigma}_*[k,k]}$ for blocks $k\in\{1,\ldots,N_E\}$.
\State Assign noise level to blocks in $\mathcal{D}^\mathcal{E}$
\NoNumber {a) Compute number of blasthole assays within each EPR block:}
\NoNumber {\qquad$c^\mathcal{E}_{*k}=\vert\{\vec{x}_i\in\mathcal{D}^\mathcal{B}\!\mid\!(a_{i,1},a_{i,2})\in[a_{*k,1}-\frac{h_{*k,1}}{2},a_{*k,1}+\frac{h_{*k,1}}{2}]\otimes [a_{*k,2}-\frac{h_{*k,2}}{2},a_{*k,2}+\frac{h_{*k,2}}{2}]\}\vert$}
\NoNumber {b) Compute average blasthole density for EPR blocks with non-zero overlap:}
\NoNumber {\qquad$\rho^\mathcal{E}=\frac{1}{M}\sum_k c^\mathcal{E}_{*k}$, where $M=\vert\{\vec{x}_{*k}\!\mid\! c^\mathcal{E}_{*k}>0\}\vert$}
\NoNumber {c) Set EPR noise level for block $\vec{x}_{*k}$, a suggested value of $\epsilon$ is 0.01}
\NoNumber {\qquad$n^{\mathcal{E}}_{k}=\begin{cases}\rho^\mathcal{E}\times\left[\left(\max_k \hat{\sigma}^\mathcal{E}(\vec{x}_{*k})\right) - \hat{\sigma}^\mathcal{E}(\vec{x}_{*k})\right] &\text{if }c^\mathcal{E}_{*k}\ne 0\\ \epsilon & \text{if }c^\mathcal{E}_{*k}=0\end{cases}$}
\NoNumber {Notes: 1) The noise level for blastholes, $n^{\mathcal{B}}_{i}$ for $i\in\{1,\ldots,N_B\}$, is assumed known.\\2) If the $c^\mathcal{E}_{*k}\neq 0$ blocks are too spatially disconnected because the blasthole sampling density is too low, counting may be expanded to include neighbouring blocks $j\!\in\!\mathcal{N}_{*k}$ in the vicinity of $\vec{x}_{*k}$. Step 5a may be modified as $c^\mathcal{E}_{*k}=\vert\{\vec{x}_i\in\mathcal{D}^\mathcal{B}\!\mid\!(a_{i,1},a_{i,2})\in\bigcup_{j\in\mathcal{N}_{*k}}[a_{*j,1}-\frac{h_{*j,1}}{2},a_{*j,1}+\frac{h_{*j,1}}{2}]\otimes [a_{*j,2}-\frac{h_{*j,2}}{2},a_{*j,2}+\frac{h_{*j,2}}{2}]\}\vert$, with the span of $\mathcal{N}_{*k}$ proportional to length scales in $\vec{\theta}$.}
\State Apply heteroscedastic GP to the combined (EPR-blasthole) dataset using the learned hyperparameters and defined EPR noise
\NoNumber {a) With $\mat{K}_y=\mat{K}+\sigma^2_y\mat{I}$ in (\ref{eq:gp-training-data-cov-matrix}), replace $\sigma^2_y\mat{I}$ with $\diag\left(\left[n^\mathcal{B}_{1},\ldots,n^\mathcal{B}_{N_B}, n^\mathcal{E}_{1},\ldots,n^\mathcal{E}_{N_E}\right]\right)^2$}
\NoNumber {b) Compute $\vec{\mu}_*$ and $\mat{\Sigma}_*$ with $\mat{X}=\left[\mat{X}^\mathcal{B}\!\mid\!\mat{X}^\mathcal{E}\right]$, $\mat{X}_*=\mat{X}^\mathcal{E}$ and $\vec{y}=\left[\vec{y}^\mathcal{B}\!\mid\!\vec{y}^\mathcal{E}\right]$.}
 \caption{--- \textbf{Data fusion using integral GP}}\label{alg:fusion}
\end{algorithmic}
\end{algorithm*}

\section{Experiment Results}\label{sec:experiment-results}
This section evaluates the proposed method using real data obtained from the Paraburdoo iron ore deposit located in Western Australia \citep{thorne2014structural}. The aim is to provide performance analysis and clarify key aspects of the data fusion algorithm. The data consists of three parts. First, the EPR block model for RL/elevation 610-620m ($\mathcal{D}^\mathcal{E}_{620}$) describes both the input and output geometry (inference grids). Second, the bench-above blasthole assay samples from 620-630m ($\mathcal{D}^\mathcal{B}_{630}$) is used for GP learning and data fusion. Finally, blasthole samples within the predicted bench from 610-620m ($\mathcal{D}^\mathcal{B}_{620}$) is withheld during training and used only for validation. The modelled region measures $480\times 255$m in the x-y plane. The blasthole samples are generally spaced 3-7m apart, and there are 1906 and 1983 samples in $\mathcal{D}^\mathcal{B}_{630}$ and $\mathcal{D}^\mathcal{B}_{620}$, respectively.

\begin{figure*}[!h]
\centering
\includegraphics{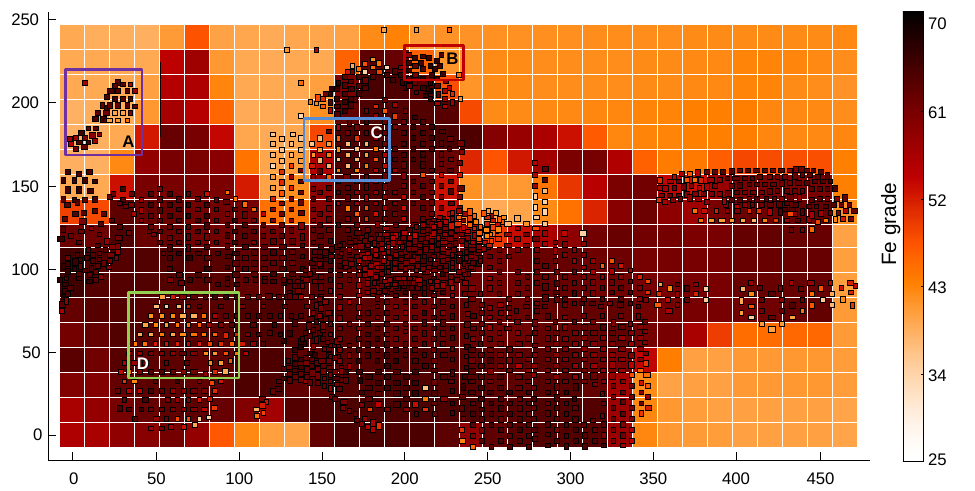}
\caption{EPR model initial Fe estimates for RL 610-620m. Bench-above blasthole assays overlaid as tiny squares.}\label{fig:analysis1-epr-init-train-blastholes}
\end{figure*}

Figure~\ref{fig:analysis1-epr-init-train-blastholes} shows a 2D cross-section of the EPR model ($\mathcal{D}^\mathcal{E}_{620}$) with the top surface at 620m. The Fe value for each EPR block is rendered in various shades of red within each $15\times 15$m tile. Likewise, the Fe interval measurements from blastholes in the bench-above ($\mathcal{D}^\mathcal{B}_{630}$) are represented by tiny squares using a common colour scale. The windows labelled A and B highlight areas of significant discrepancies in which high grade iron is misrepresented as low grade in the EPR blocks. The opposite situation is highlighted in windows C and D where low grade iron is misrepresented as high grade ore in the EPR model. These differences have real implications for grade control in mining operations. Ore dilution may occur when a significant quantity of low grade or waste material is mistaken as high grade ore. These issues are precisely what the data fusion algorithm is motivated to address and rectify.

\subsection{Data assimilation}\label{sec:analysis-assimilation}
\begin{figure*}[!h]
\centering
\includegraphics{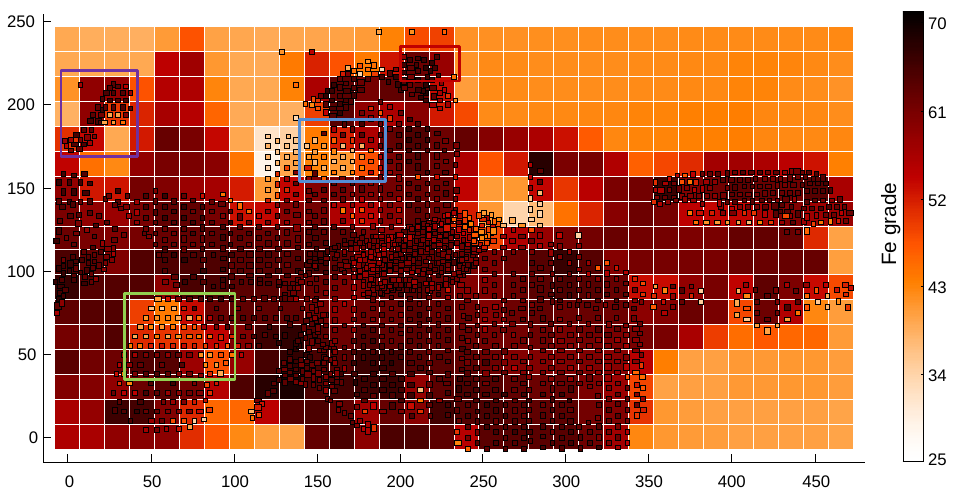}
\caption{EPR+BH GP fusion mean Fe estimates for RL 610-620m. Bench-above blasthole assays overlaid as tiny squares.}\label{fig:analysis1-epr-gpmf-train-blastholes}
\end{figure*}

Figure~\ref{fig:analysis1-epr-gpmf-train-blastholes} shows the qualitative improvement when the bench-above blasthole assays ($\mathcal{D}^\mathcal{B}_{630}$) are fused with the rudimentary EPR block model ($\mathcal{D}^\mathcal{E}_{620}$). In the highlighted areas, the colour of the EPR blocks (hence, the modelled Fe grade) have now become consistent with the enclosed blasthole samples. Before delving into detailed analysis, it is worth emphasising the mechanism which enables this to be achieved.

\begin{figure*}[!h]
\centering
\includegraphics{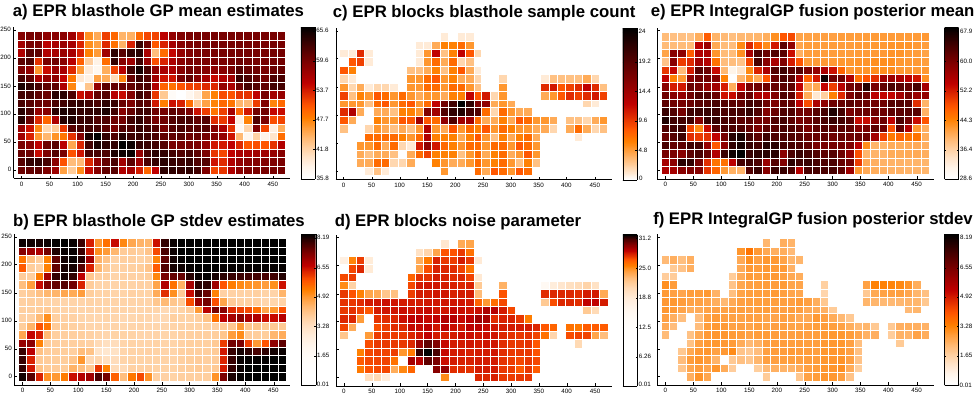}
\caption{Understanding the noise tradeoff between the EPR data and bench-above blasthole assays.}\label{fig:analysis1-noise-illustrated}
\end{figure*}

The optimal hyperparameters $\vec{\theta}=[\sigma^2_f,l_x,l_y,l_z,n^\mathcal{B}]=[8.863,20.593,15.052,38.756,2.321]$ were obtained from Step 3 in Algorithm~\ref{alg:fusion} by applying GP learning to $\mathcal{D}^\mathcal{B}_{630}$. Subsequently, Fig.~\ref{fig:analysis1-noise-illustrated} (a)--(b) show the GP mean and standard deviation estimates, $\vec{\mu}_*$ and $\hat{\vec{\sigma}}^\mathcal{E}$, computed for EPR blocks, $\vec{x}_*$ from 610-620m, using only the training blastholes from 620-630m. As is a common feature of GP, the prediction trends toward the mean in the NE and SE corners of Fig.~\ref{fig:analysis1-noise-illustrated}(a) and near the margins more generally, as the estimated locations get further from the known data (blasthole measurements). The dark patches in Fig.~\ref{fig:analysis1-noise-illustrated}(b) indicate areas with high uncertainty, where on its own, GP regression using $\mathcal{D}^\mathcal{B}_{630}$ would yield neither confident nor accurate predictions. Conversely, the light patches in Fig.~\ref{fig:analysis1-noise-illustrated}(b) correspond to areas where blasthole information can be leveraged to improve the initial Fe values in the EPR model. Figure~\ref{fig:analysis1-noise-illustrated}(c) counts the number of blasthole assays within each EPR block ($c^\mathcal{E}_{*k}$), this is used to compute the EPR noise level ($n^\mathcal{E}_k$) in Step 5c in Algorithm~\ref{alg:fusion}. The result in Fig.~\ref{fig:analysis1-noise-illustrated}(d) shows the noise parameter is amplified where blasthole evidence is plentiful to suppress the contribution of the EPR data. This allows the final Fe prediction to gravitate towards blasthole measurements particularly when the blasthole-induced uncertainty, $\hat{\sigma}^\mathcal{E}(\vec{x}_{*k})$, is low. As seen previously in Fig.~\ref{fig:analysis1-epr-gpmf-train-blastholes}, this promotes consistency between the EPR and training blasthole data. By the same token, the IntegralGP fusion algorithm defers the regression to the EPR model when the blasthole prediction uncertainty, $\hat{\sigma}^\mathcal{E}(\vec{x}_{*k})$, is high. In this case, the noise parameter ($n^\mathcal{E}_k$) is set close to zero not because the EPR model has perfect knowledge, or the block value is deemed to be error-free, this initial EPR value is preserved rather than overturned---when there is insufficient evidence from the alternative source---to respect the authority and experience of experts who have shaped the EPR model. In Fig.~\ref{fig:analysis1-noise-illustrated}(e), the IntegralGP fusion posterior mean distribution shows the influence of the EPR model, which is responsible for reverting dark patches to light patches in the NE and SE corners. Figure~\ref{fig:analysis1-noise-illustrated}(f) shows how the deficiencies of the blasthole source is addressed by data fusion; it reduces the uncertainty in high $\hat{\sigma}^\mathcal{E}(\vec{x}_{*k})$ regions where assay samples are unavailable.

\subsection{Model validation}\label{sec:validate}
\begin{figure*}[!h]
\centering
\includegraphics{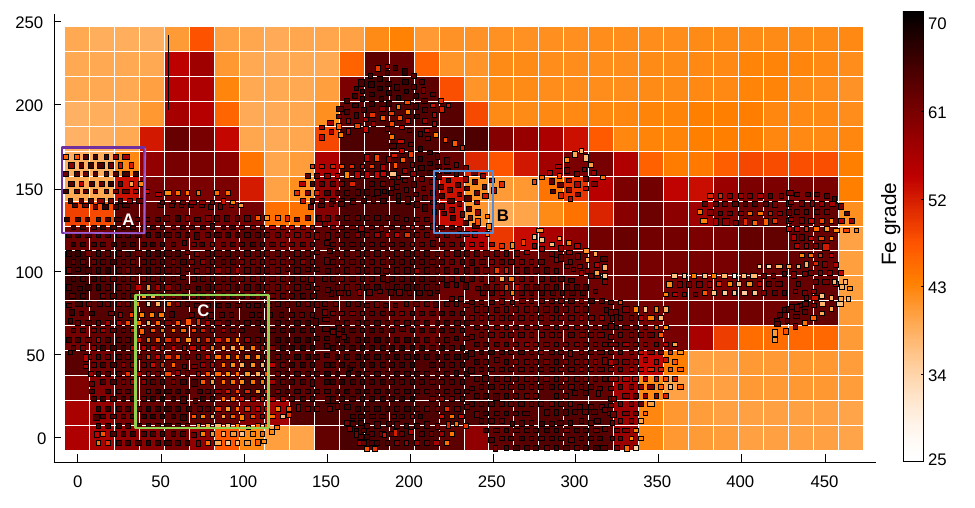}
\caption{EPR model initial Fe estimates for RL 610-620m. In-situ validation blastholes (within the predicted bench) are shown as tiny squares.}\label{fig:analysis1-epr-init-validation-blastholes}
\end{figure*}

\begin{figure*}[!h]
\centering
\includegraphics{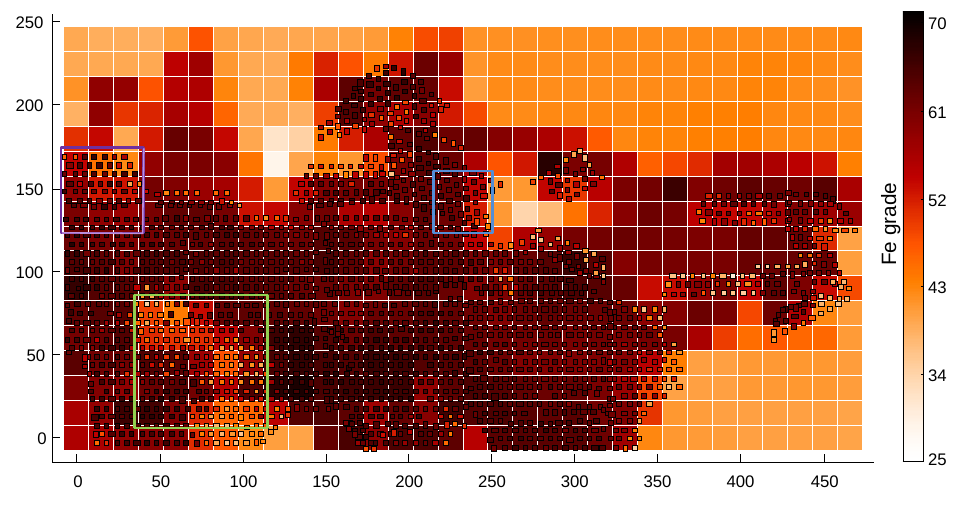}
\caption{EPR+BH GP fusion mean Fe estimates for RL 610-620m. In-situ validation blastholes (within the predicted bench) are shown as tiny squares.}\label{fig:analysis1-epr-gpmf-validation-blastholes}
\end{figure*}

The analysis thus far has focused on model comparison with respect to $\mathcal{D}^\mathcal{B}_{630}$, the bench-above blastholes. While this gives good insight into how data assimilation works, the test for predictive performance ought to be made against $\mathcal{D}^\mathcal{B}_{620}$, the in-situ validation blasthole assays taken within the predicted bench. Following earlier conventions, the EPR model before and after data fusion are compared with $\mathcal{D}^\mathcal{B}_{620}$ in Figs.~\ref{fig:analysis1-epr-init-validation-blastholes} and \ref{fig:analysis1-epr-gpmf-validation-blastholes}. Windows A to C show the IntegralGP fusion model also reconciles better with validation blastholes within 610-620m.

\begin{figure*}[!h]
\centering
\includegraphics{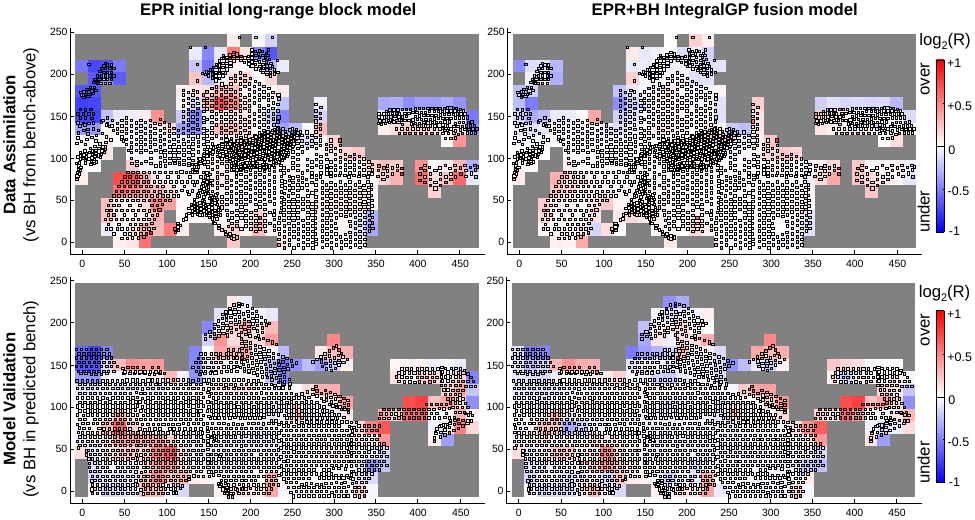}
\caption{Comparison of the initial EPR estimates with the EPR/BH IntegralGP fusion model constructed for RL 610-620m. Top row: using bench-above blastholes $\mathcal{D}^\mathcal{B}_{630}$ to emphasise data assimilation. Bottom row: using bench-within blastholes $\mathcal{D}^\mathcal{B}_{620}$ for model validation. These results correspond to Figs.~\ref{fig:analysis1-epr-init-train-blastholes}, \ref{fig:analysis1-epr-gpmf-train-blastholes}, \ref{fig:analysis1-epr-init-validation-blastholes} and \ref{fig:analysis1-epr-gpmf-validation-blastholes}.}\label{fig:analysis2-ratios-epr-init-gpmf-vs-blastholes-tv}
\end{figure*}

To illustrate these differences more clearly, a red-blue colour palette is used in Fig.~\ref{fig:analysis2-ratios-epr-init-gpmf-vs-blastholes-tv} to highlight the magnitude and areas of over- and under-estimation. What is plotted is the log-ratio of $R(\vec{x}_*)=\frac{\hat{\mu}(\vec{x}_*)}{y_0(\vec{x}_*)}$, between the mean model predictions and true Fe values informed by the blasthole samples. The top row uses the bench-above blasthole assays $\mathcal{D}^\mathcal{B}_{630}$ as reference to emphasise the efficacy of blasthole fusion. The bottom row uses a different reference, viz., the bench-within blasthole assays $\mathcal{D}^\mathcal{B}_{620}$ to emphasise predictive performance. Visually, the reduction in colour intensity in the top-right panel shows the available blasthole information has been incorporated successfully into the fusion model. Furthermore, the bottom-right panel demonstrates that predictive performance has also improved following IntegralGP fusion. Key differences can be seen in the west and south-west in Fig.~\ref{fig:analysis2-ratios-epr-init-gpmf-vs-blastholes-tv}. To quantify these changes, the distortion measure $\hat{\sigma}_R$ defined in (\ref{eq:sigmaR}) is employed. Since this involves a ratio, the standard deviation is computed in the logarithmic domain (with $g(r)=\log_2(r), g'(r)=\frac{1}{\text{ln}(2)\cdot r}$) and transformed using Taylor series approximation.\footnote{The variance relates to the random function via $\text{var}[R]\approx\text{var}[g(R)] / (g'(\mu_R))^2$, where $\mu_R=\mathbb{E}[R]=1$}
\begin{equation}
\hat{\sigma}_R\approx\sqrt{\text{var}[\log_2(R)]}\cdot\text{ln}(2)\label{eq:sigmaR}
\end{equation}

\begin{table*}[!h]
\centering
\resizebox{0.9\textwidth}{!}{
\begin{tabular}{l|cc}\hline
& \multicolumn{2}{c}{\small Model}\\
{\small Reference blasthole assays} & EPR initial Fe estimates & EPR+BH IntegralGP fusion\\\hline
{\small Bench-above $\mathcal{D}^\mathcal{B}_{630}$} & 0.165 & 0.074\\
{\small Bench-within $\mathcal{D}^\mathcal{B}_{620}$} & 0.151 & 0.119\\\hline
\end{tabular}
}
\caption{Model predictive performance as measured by the distortion metric $\hat{\sigma}_R$ with respect to a set of reference blasthole assays. These results correspond to Fig.~\ref{fig:analysis2-ratios-epr-init-gpmf-vs-blastholes-tv}.}\label{tab:sigmaR}
\end{table*}
The numerical results presented in Table~\ref{tab:sigmaR} confirm the improvements observed with IntegralGP fusion from the perspective of data assimilation (in-situ regression) and bench-below prediction (extrapolation to 610-620m) using only blastholes from 620-630m. This is a significant result as it shows the blasthole assays (interval measurements) gathered from 620-630m can be utilised to improve prediction accuracy at 610-620m. Moreover, this data is plentiful and may be collected during mining production following standard procedures.

\subsection{Material classification}\label{sec:material-classification}
In this section, we apply categorical judgement to analyse and reinterpret the results in a manner that is more closely aligned with current industry practice. In surface mining, the material excavated from the pit are transported by haul trucks to different destinations to maintain a steady supply of material \citep{seiler2020flow} to satisfy flow constraints and ore blending requirements downstream. In Pilbara mines, iron ore is often classified into three categories as follows \citep{sommerville2014mineral} based on the Fe grade.
\begin{align}
\text{Category:}\quad c(\vec{x}_*)&=\begin{cases}1\quad\text{for waste (W)} & \hat{\mu}_\text{Fe}(\vec{x}_*)<55\\2\quad\text{for low-grade (LG)} & 55\le\hat{\mu}_\text{Fe}(\vec{x}_*)<60\\3\quad\text{for high-grade (HG)} & \hat{\mu}_\text{Fe}(\vec{x}_*)\ge 60\end{cases}\label{eq:category-thresholds}
\end{align}
\noindent Depending on the classification, this would allow high-grade ore to be moved to a crusher for instance, low-grade material to be transferred to an appropriate stockpile, and unrecoverable material to be disposed at a waste dump.

\begin{figure*}[!h]
\centering
\includegraphics{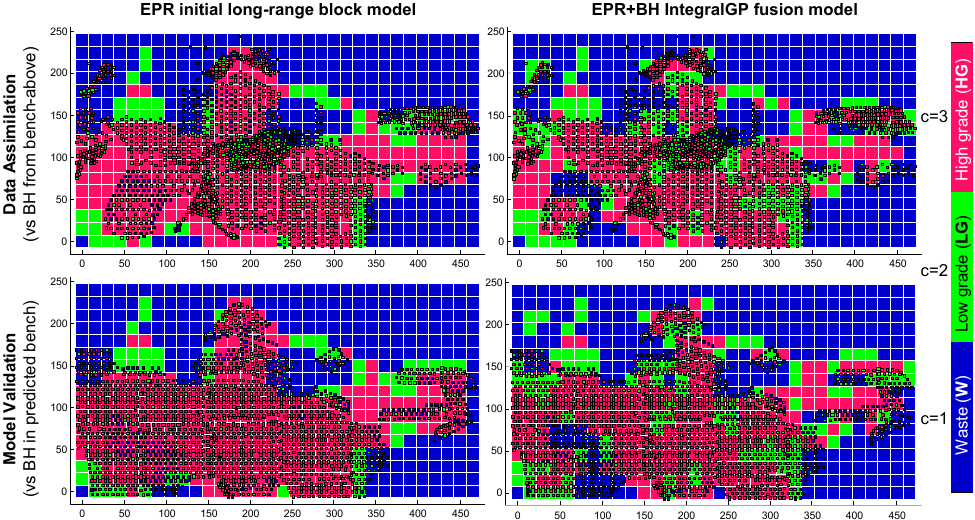}
\caption{Material classification using the initial EPR estimates and EPR/BH IntegralGP fusion model for RL 610-620m. These results correspond to Figs.~\ref{fig:analysis1-epr-init-train-blastholes}, \ref{fig:analysis1-epr-gpmf-train-blastholes}, \ref{fig:analysis1-epr-init-validation-blastholes} and \ref{fig:analysis1-epr-gpmf-validation-blastholes}.}\label{fig:analysis3-categories-epr-init-gpmf-vs-blastholes-tv}
\end{figure*}

\begin{figure*}[!h]
\centering
\includegraphics{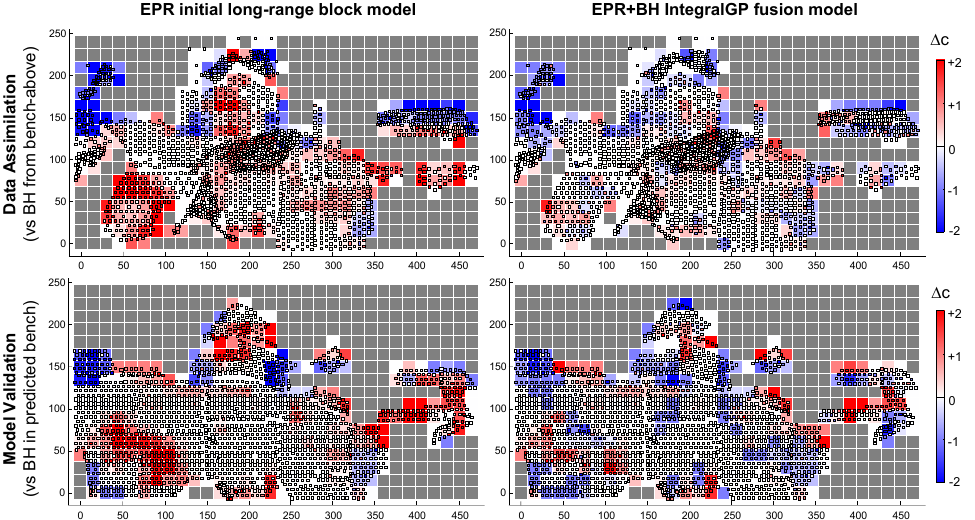}
\caption{Categorical distance between model and reference for material classification using the initial EPR estimates and EPR/BH IntegralGP fusion model.}\label{fig:analysis3-magnitude-of-confusion-epr-init-gpmf-vs-blastholes-tv}
\end{figure*}

Figure~\ref{fig:analysis3-categories-epr-init-gpmf-vs-blastholes-tv} shows the material classification based on the initial EPR estimates, and EPR/BH IntegralGP fusion model. As before, two different comparisons are made. Bench-above blastholes $\mathcal{D}^\mathcal{B}_{630}$ serve as the reference in the top row to highlight the benefits of data fusion for in-situ regression. Bench-within (validation) blastholes $\mathcal{D}^\mathcal{B}_{620}$ are used in the bottom row to assess predictive performance. It can be shown the improvements previously observed in the IntegralGP fusion model also hold following discretisation. However, these differences may not be apparent from the graphics in their current form. To reveal these differences and facilitate quantitative analysis, the concept of categorical distance is defined in (\ref{eq:categorical-distance}) to emphasise the magnitude of error in a categorical prediction. This is based on the intuition that classifying a LG block as HG is not as detrimental as treating a W (waste) block as HG.
\begin{equation}
\text{Categorical distance:}\quad \Delta c(\vec{x}_*)=c_\text{model}(\vec{x}_*) - c_\text{reference}(\vec{x}_*)\in[-2,2]\label{eq:categorical-distance}
\end{equation}
\noindent 
Inspection of Fig.~\ref{fig:analysis3-magnitude-of-confusion-epr-init-gpmf-vs-blastholes-tv} shows the colours for IntegralGP fusion are on average less intense. This corroborates the earlier claim that IntegralGP fusion produces similar benefits even with Fe grade quantisation. To verify this, the average and mean absolute categorical distance, mean\,$\lvert\Delta c\rvert$ and $\mathbb{E}\left[\Delta c\right]$, are computed for the EPR initial estimates and EPR/BH IntegralGP fusion model. The metrics reported in Table~\ref{tab:cat-performance} demonstrate that IntegralGP fusion minimises the categorical prediction error, and significantly reduces the model bias. This provides a practical measure of overall classification performance that emphasises the varying levels of confusion ($\Delta c$) in the prediction outcome. The conditional probability $p(\hat{C}=\text{HG}\!\mid\!C_0=\text{W})$ in Table~\ref{tab:cat-cond-prob} also shows the likelihood of incorrectly classifying waste as high-grade is significantly reduced by the IntegralGP fusion model.

\begin{table*}[!h]
\centering
\resizebox{0.95\textwidth}{!}{
\begin{tabular}{l|c|cc}\hline
& & \multicolumn{2}{c}{\small Model}\\
{\small Reference blasthole assays} & Metric & EPR initial Fe estimates & EPR+BH IntegralGP fusion\\\hline
{\small Bench-above $\mathcal{D}^\mathcal{B}_{630}$} & {\small mean\,$\lvert\Delta c\rvert$} & 0.616 & 0.360\\
& $\mathbb{E}\left[\Delta c\right]$ & 0.157 & -0.085\\\hline
{\small Bench-within $\mathcal{D}^\mathcal{B}_{620}$} & {\small mean\,$\lvert\Delta c\rvert$} & 0.534 & 0.469\\
& $\mathbb{E}\left[\Delta c\right]$ & 0.247 & -0.026\\\hline
\end{tabular}
}
\caption{Categorical prediction model performance metrics: $\text{mean}\,\lvert\Delta c\rvert$ indicates the magnitude of error, $\mathbb{E}[\Delta c]$ indicates model bias.}\label{tab:cat-performance}
\end{table*}

\begin{table*}[!h]
\centering\small
\begin{tabular}{rr|p{10mm}p{10mm}p{10mm}p{10mm}|p{10mm}p{10mm}p{10mm}p{10mm}}\hline
& & \multicolumn{4}{c|}{EPR initial long-range model} & \multicolumn{4}{c}{EPR/BH IntegralGP fusion}\\
& $C_0$ & \multicolumn{4}{c|}{predicted class, $\hat{C}$} & \multicolumn{4}{c}{predicted class, $\hat{C}$}\\\hline
\parbox[t]{2mm}{\multirow{5}{*}{\rotatebox[origin=c]{90}{\small $\mathcal{D}^\mathcal{B}_{630}$ BH}}} & & W & LG & HG & $n(C_0)$ & W & LG & HG & $n(C_0)$\\
& W & 0.389 & 0.105 & \underline{0.506} & 342 & 0.707 & 0.205 & \underline{0.088} & 342\\
& LG & 0.151 & 0.096 & 0.753 & 429 & 0.163 & 0.389 & 0.448 & 429\\
& HG & 0.103 & 0.072 & 0.825 & 1135 & 0.053 & 0.183 & 0.764 & 1135\\
& $n(\hat{C})$ & 315 & 159 & 1432 & & 372 & 445 & 1089 & \\\hline
\parbox[t]{2mm}{\multirow{5}{*}{\rotatebox[origin=c]{90}{$\mathcal{D}^\mathcal{B}_{620}$ BH}}} & & W & LG & HG & $n(C_0)$ & W & LG & HG & $n(C_0)$\\
& W & 0.222 & 0.131 & \underline{0.647} & 414 & 0.526 & 0.196 & \underline{0.280} & 414\\
& LG & 0.169 & 0.128 & 0.703 & 219 & 0.251 &  0.288 &  0.461 & 219\\
& HG & 0.057 & 0.058 & 0.885 & 1350 & 0.060 & 0.217 & 0.723 & 1350\\
& $n(\hat{C})$ & 206 & 160 & 1617 & & 354 & 436 & 1193 & \\\hline
\end{tabular}
\caption{Categorical prediction conditional probabillities $p(\hat{C}\!\mid\!C_0)$ and sample counts---$n(\hat{C})$ for predicted class and $n(C_0)$ for actual class. $\mathcal{D}^\mathcal{B}_{630}$ and $\mathcal{D}^\mathcal{B}_{620}$ denote reference and groundtruth established by bench-above blasthole assays (620-630m) and those within the predicted bench (610-620m).}\label{tab:cat-cond-prob}
\end{table*}

\subsection{Ablative analysis}\label{sec:ablative}
\begin{figure*}[!h]
\centering
\includegraphics{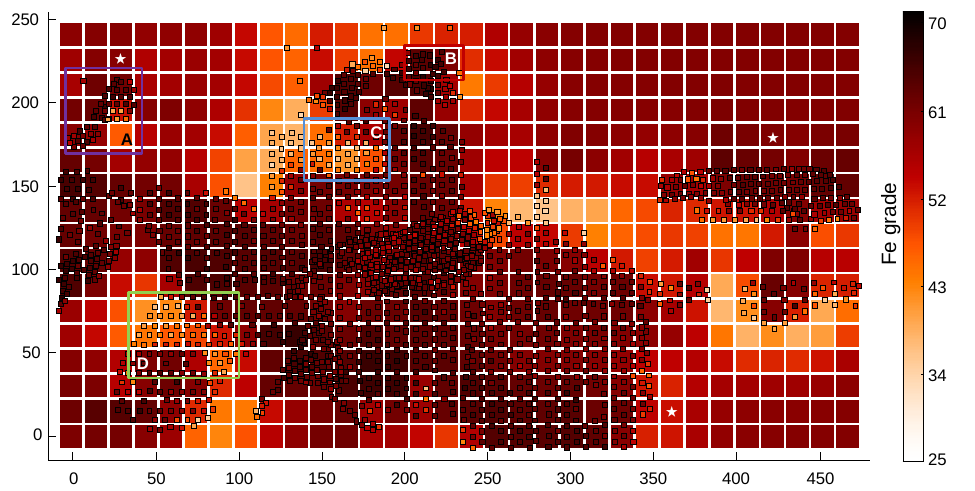}
\caption{BH-only IntegralGP mean Fe estimates for RL 610-620m. Bench-above blasthole assays overlaid as tiny squares. Corresponding results for EPR initial model and EPR+BH fusion appear in Figs.~\ref{fig:analysis1-epr-init-train-blastholes} and \ref{fig:analysis1-epr-gpmf-train-blastholes}}\label{fig:analysis1-bh-only-gpmf-train-blastholes}
\end{figure*}

To gain a deeper understanding of the system's behaviour without data fusion, the EPR source shall now be removed. This results in a BH-only IntegralGP grade prediction model that uses only the bench-above blasthole assays ($\mathcal{D}^{\mathcal{B}}_{630}$) without the prior knowledge contained in $\mathcal{D}^{\mathcal{E}}_{620}$. The GP mean estimates for the BH-only model are shown in Fig.~\ref{fig:analysis1-bh-only-gpmf-train-blastholes}. Compared with the EPR initial model, the discrepancies between the block estimates and BH assays seen in Fig.~\ref{fig:analysis1-epr-init-train-blastholes} are largely eliminated in the BH-only model, just as it did in the EPR+BH fusion model in Fig.~\ref{fig:analysis1-epr-gpmf-train-blastholes}. However, in regions that lack BH training data (such as the northeast and southeast corners of Fig.~\ref{fig:analysis1-bh-only-gpmf-train-blastholes}), the BH-only model predictions regress toward the mean; this results in gross over-estimation of the grade which is a significant problem. Furthermore, the GP prediction is highly uncertain in data sparse regions as highlighted in Fig.~\ref{fig:analysis1-noise-illustrated}b. The contrast is visually striking, considering for instance, the block estimates at the $\star$ locations are red in the BH-only model (Fig.~\ref{fig:analysis1-bh-only-gpmf-train-blastholes}), but orange in the EPR+BH fusion model (Fig.~\ref{fig:analysis1-epr-gpmf-train-blastholes}). This phenomenon, the tendency of the GP mean drifting back to the dataset average in data deficient areas, was pointed out by \citet{ball2021geostat}. It reflects the reality that blastholes are more densely drilled in ore-rich regions---particularly near complex boundaries---and there is limited incentive for sampling in regions that are thought to be unmineralised, where the excavated material would most likely be transferred to a waste dump. This creates a situation of class imbalance where high-grade samples are over-represented. It emphasises the importance of integrating expert knowledge of the orebody, as in the proposed fusion algorithm, or constraining the solution to produce geologically plausible estimates.

\begin{figure*}[!h]
\centering
\includegraphics{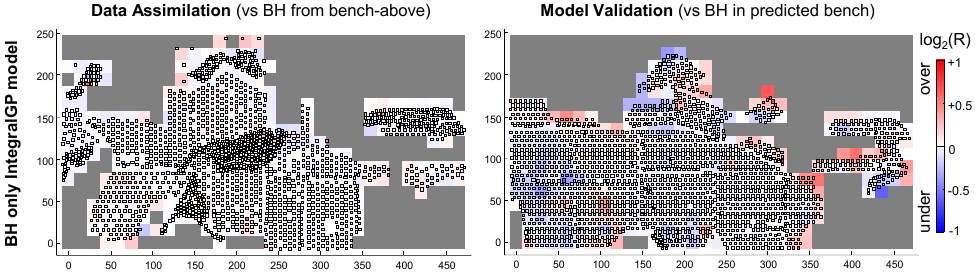}
\caption{BH-only GP model predictions: depicting ratio of estimates to assay values in RL 610-620m. Left: referencing bench-above blastholes $\mathcal{D}^\mathcal{B}_{630}$ to emphasise data assimilation. Right: referencing bench-within blastholes $\mathcal{D}^\mathcal{B}_{620}$ for model validation. These results correspond to Fig.~\ref{fig:analysis2-ratios-epr-init-gpmf-vs-blastholes-tv}.}\label{fig:analysis2-ratios-bh-only-gpmf-vs-blastholes-tv}
\end{figure*}

\begin{figure*}[!h]
\centering
\includegraphics{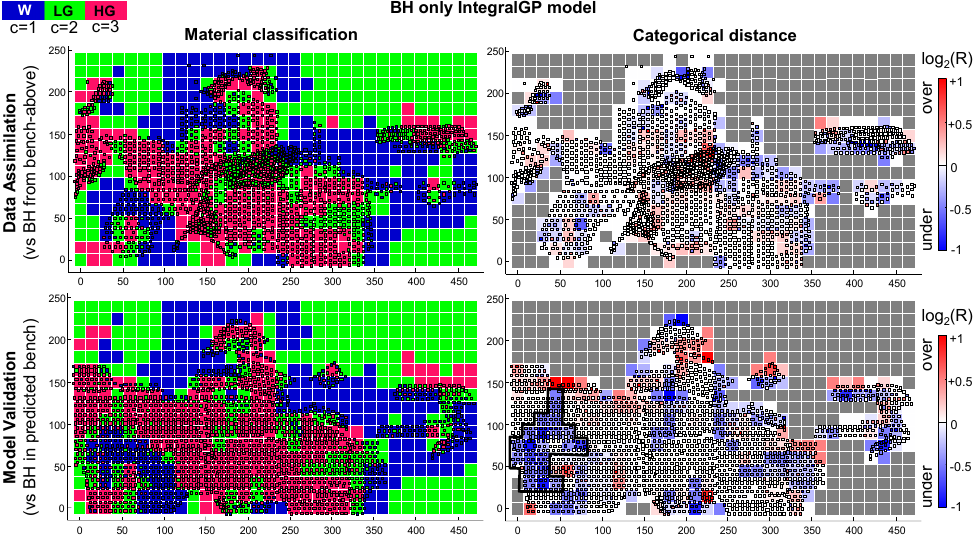}
\caption{Material classification and categorical distance for the BH-only GP model in RL 610-620m. Top row: referencing bench-above blastholes $\mathcal{D}^\mathcal{B}_{630}$ to emphasise data assimilation. Bottom row: referencing bench-within blastholes $\mathcal{D}^\mathcal{B}_{620}$ for model validation. These results correspond to Figs.~\ref{fig:analysis3-categories-epr-init-gpmf-vs-blastholes-tv} and \ref{fig:analysis3-magnitude-of-confusion-epr-init-gpmf-vs-blastholes-tv}.}\label{fig:analysis3-categories-bh-only-gpmf-vs-blastholes-tv}
\end{figure*}

Figure~\ref{fig:analysis2-ratios-bh-only-gpmf-vs-blastholes-tv} compares the BH-only model with assays for the prediction interval RL 610-620m. Referencing the bench-above blastholes reveals stronger adherence to the observed blasthole values in the case of BH-only which is to be expected; the computed distortion for BH-only and EPR+BH fusion are 0.033 and 0.074, respectively. However, the predicted grades are not demonstrably better when validated against bench-within blastholes, the distortion for BH-only and EPR+BH fusion are comparable (0.100 and 0.119, respectively, which are nonetheless lower than 0.151 for the initial EPR model). An important limitation that must be acknowledged is that large sections (particularly low grade and waste zones) are not covered by validation blastholes ($\mathcal{D}^{\mathcal{B}}_{620}$). Thus, the errors for the BH-only model are significantly under-reported. The green blocks in the northwest, northeast and southeast corners of Fig.~\ref{fig:analysis3-categories-bh-only-gpmf-vs-blastholes-tv}(left) serve as a reminder of this. For material classification, the categorical distance was computed for the BH-only model, its mean absolute error, $\text{mean}\lvert\Delta c\rvert$, is 0.413 compared with 0.469 for EPR+BH fusion. Once again, this masquerades the inferior performance of the BH-only model in data deficient regions. However, a noticeable negative bias (absent from the EPR+BH fusion model) is revealed in the BH-only model, $\mathbb{E}[\Delta c]$ is -0.131 and -0.026, respectively, for BH-only and EPR+BH fusion. This is evident from the blue cluster demarcated by the black line in Fig.~\ref{fig:analysis3-categories-bh-only-gpmf-vs-blastholes-tv}(bottom-right), which gives only a glimpse of the problem of GP extrapolation in areas devoid of BH data when EPR fusion is completely disabled.

\subsection{Discussion}\label{sec:discuss}
The proceeding analysis demonstrates that our proposal can improve geochemical grade modelling in the context of bench-below prediction in an open-pit mine. While the results are encouraging, the precision and recall rates are far from perfect. This is to be expected since the experiments were designed to focus on data fusion rather than other confounding variables. If wireframes were supplied for boundary delineation, the modelling space could be partitioned into multiple domains to make geochemical characteristics more homogeneous within each region. This would allow a separate GP to be constructed for each domain to effectively model a unimodal distribution rather than a multimodal one. These approaches are described in \citep{lowe2021bayesian,balamurali2022surface}. The orientation of an orebody can also affect regression performance if the spatial correlations captured by the kernels are not aligned with stratigraphic structures. To address this, the data may be rotated---with expert guidance or unsupervised---to encourage information to propagate in the right direction between successive benches \citep{ball2022creating}.
\begin{figure*}[!h]
\centering
\includegraphics{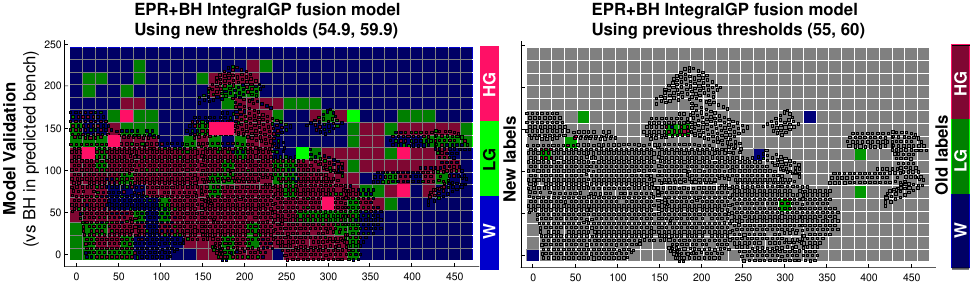}
\caption{Material classification---sensitivity to decision thresholds. Left: Using (54.9, 59.9) as the thresholds in (\ref{eq:category-thresholds}). Existing labels are darken. Changed labels appear brighter. Right: Using (55, 60) as the thresholds. Only the previous classification for changed labels are shown.}\label{fig:analysis3-thresholds-sensitivity}
\end{figure*}

\begin{figure*}[!h]
\centering
\includegraphics{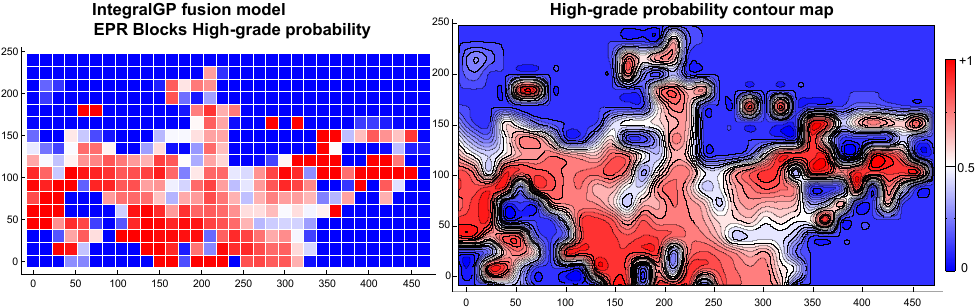}
\caption{Probabilistic prediction for HG class. Left: $p(C(\vec{x}_*)=\text{HG}\!\mid\!\hat{\mu}(\vec{x}_*),\hat{\sigma}(\vec{x}_*))$. Right: Level-set representation of HG likelihood.}\label{fig:analysis4-epr-gpmf-hg-pr-contours}
\end{figure*}

Material classification can be very sensitive to the decision thresholds especially when grade uncertainty is disregarded. Figure~\ref{fig:analysis3-thresholds-sensitivity} shows substantial changes can occur even when the thresholds are slightly perturbed. As mining operations become increasingly complex, probabilistic predictions provide a foundation for handling uncertainty in a robust way. For instance, the likelihood of encountering HG material may be expressed as $p(C(\vec{x}_*)=\text{HG}\!\mid\! \hat{\mu}(\vec{x}_*),\hat{\sigma}(\vec{x}_*)) = 1 - \Phi\left(Z<(60-\hat{\mu})/\hat{\sigma}\right)$, where $\Phi(z)$ represents the CDF of the standard normal distribution. This allows mine planners to take a stochastic approach to grade control and material excavation; develop, evaluate or modify policies to mitigate adverse events such as ore dilution according to engineering requirements and risk appetite. Within the expert system, the movements of diggers and haul trucks can be displayed in a probabilistic material classification map (see Fig.~\ref{fig:analysis4-epr-gpmf-hg-pr-contours}) to increase situational awareness and facilitate precision mining.

\newpage
To maintain a coherent narrative, some finer details and practicalities have been omitted from this discussion. For completeness, {\color{ruby}Appendix~\ref{app:research-perspective}} provides a synopsis of recent work and an account of data fusion from a mining perspective. Readers are referred to {\color{ruby}Appendix~\ref{app:practicalities}} for other considerations relating to kernel anisotropy, hyperparameters optimisation and implementation.

\subsubsection{Future work}\label{sec:future-work}
The examples in this paper are focused on fusion of point, interval and block data, however the proposed methodology also allows incorporation of data from surface areas. Future work may include fusion of data from mine wall and mine floor surfaces collected through hyperspectral imagery \citep{windrim2023unsupervised}, lab analysis and manual inspection. Fusion of surface data with other information sources (such as from exploration and blast holes) will enable better modelling of the material properties behind the walls and below the floors which can be instrumental for improved bench below and cutback mining.

\section{Conclusions}\label{sec:conclusion}
In this paper, we presented an IntegralGP framework for volumetric estimation of subterranean geochemical properties in a mineral deposit. At a high level, it relates to the inference engine in a mining-oriented expert system, where computational intelligence is concerned with knowledge representation, uncertainty estimation, probabilistic reasoning and using machine learning to predict target attributes at unobserved locations. This expert system can also be portrayed as an autonomous system that provides data fusion and model update capability. Specifically, its objective is to provide useful grade estimates beneath the drilled bench. This scenario is depicted in Fig.~\ref{fig:pit-benches}. The challenges, or constraints, come in the form of sparse data with variable resolution and sample spacing; the system must utilise and improve predictions of the long-range block model---that is, rectifying local inaccuracies while respecting the general structure and geochemical distribution specified by domain experts. Driven by practical considerations and the top-down manner in which open-pit mining operates, the desirable outcomes would be an increased ability to predict the spatial distribution of minerals (e.g. Fe concentration) in the bench below, and boundary delineation to separate high-grade, low-grade or waste material. As motivation, Sec.~\ref{sec:motivation} demonstrates that IntegralGP enables both these outcomes---more accurate regression and boundary localisation---to be achieved when the different spatial supports of measurements were properly considered.

IntegralGP is built upon regular Gaussian Process, a nonparametric supervised learning technique that is used to solve probabilistic regression and classification problems. In our formulation, heterogeneous data---for instance, block data and interval measurements which are chemically correlated but spatially incongruous---are represented in a unified way in the observation model, whereby incorporating their spatial dimensions basically allows the standard GP training and inference procedures to be left intact. What is different is how the covariance terms (for instance $\mat{K}_{*}[i,j]=k(\vec{x}_{i},\vec{x}_{*j})$) are computed; these appear in the expressions of the GP posterior mean ($\vec{\mu}_*$) and covariance matrix ($\mat{\Sigma}_*$). Using separable kernels, a general expression accounting for different combinations of spatial support, such as line-volume covariance, was presented in (\ref{eq:cov-yi-uj}). To facilitate hyperparameters optimisation, covariance partial derivatives with respect to length-scale parameters were also obtained. Concrete expressions for several families including the Mat\'ern 3/2 kernel ($\varphi$) and its anti-derivatives ($\Phi$ and $\Psi$) are reported in Tables~\ref{tab:varphi-grad} and \ref{tab:Phi-Psi}. Numerical stability issues associated with large distances were discussed. These covariance expressions developed for data with different spatial supports underpin IntegralGP and constitute our first contribution.

To illustrate its application, a fusion algorithm was described drawing on the principle of heteroscedastic GP regression. The objective was to show how blasthole assays from the bench-above (620-630m) can be utilised to improve the predictive performance of the initial EPR block model in estimating the Fe composition in a future bench (610-620m). Using visualisation, the data assimilation and noise tradeoff features of the algorithm were explained. Blasthole assays from the 610-620 bench were used for model validation. Using the model-vs-groundtruth ratio standard deviation, $\hat{\sigma}_R$, it was shown the EPR/BH IntegralGP fusion model yields better regression results compared with the initial EPR long-range block model. Subsequently, the predicted Fe grade was discretised consistent with current industry practice and classified into three categories: waste (W), low-grade (LG) or high-grade (HG) using fixed decision thresholds. Analysis based on the categorical distance, $\Delta c$, revealed EPR/BH fusion can minimise the absolute error and reduce the bias in categorical prediction. In particular, the likelihood of misclassifying W as HG was significantly reduced. However, the classification results are sensitive to changes in the decision thresholds. To address this, we showed how the GP mean and variance estimates can be used jointly (as intended) to perform risk-based probabilistic assessment, and help mine planners and equipment operators make more informed decisions. The data fusion algorithm and probabilistic perspective form the second part of this paper's contribution.

\renewcommand{\thesection}{A.\arabic{section}}
\renewcommand{\thefigure}{A\arabic{figure}}
\renewcommand{\thetable}{A\arabic{table}}
\renewcommand{\theequation}{A\arabic{equation}}
\setcounter{section}{0}
\setcounter{figure}{0}
\setcounter{table}{0}
\setcounter{equation}{0}

\section{Appendix A: Synthesis steps for the motivating examples}\label{app:construction}
\begin{figure}[h]
\centering
\includegraphics[width=125mm]{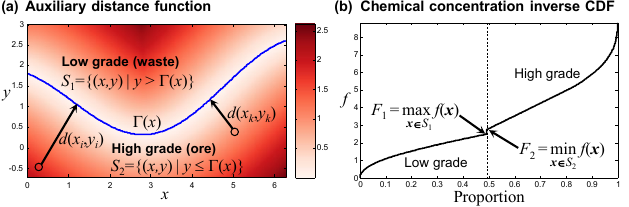}
\caption{Main elements used in the construction of the motivating examples.}\label{fig:construction}
\end{figure}

This appendix provides details on how the motivating examples in Sec.~\ref{sec:motivation} were constructed. Let $\Gamma(x)$ represents the boundary featured in Figs.~\ref{fig:motivation1} and \ref{fig:motivation2}. The boundary groundtruths are defined by $\Gamma_1(x)=1+\sin(0.1135 x)+\cos(0.85x+\frac{\pi}{5})$ and $\Gamma_2(x)=\Gamma_1(2x)$ respectively over the domain $x\in[0,2\pi]$. This boundary partitions the modelling space $\mathcal{R}_{xy}=[0,2\pi]\otimes [-0.7,3]$ into two disjoint regions that represent the low grade (waste) and high grade (ore) zones. These regions are described by $S_1=\{(x_i,y_i)\mid y_i\!>\!\Gamma(x_i)\}$ and $S_2=\{(x_i,y_i)\mid y_i\!\le\!\Gamma(x_i)\}$, respectively, as shown in Fig.~\ref{fig:construction}(a). An auxiliary function $d(x,y)$ which measures the minimum Euclidean distance from each point to the boundary is defined, this distance map is visualised in Fig.\ref{fig:construction}(a). 
To simulate the chemical spatial distribution, the formula in (\ref{eq:target-attribute-synthesis}) is applied to produce the target functions shown in Figs.~\ref{fig:motivation1}a and \ref{fig:motivation2}a.
\begin{align}
f(x,y)&=\begin{cases}d_{\max} - d(x,y) & \text{if }(x,y)\in S_1\\
\kappa\times d(x,y) + d_{\max} +\delta & \text{if }(x,y)\in S_2\end{cases}\label{eq:target-attribute-synthesis}\\
\text{where }& d_{\max}=\max_{(x,y)\in S_1}d(x,y), \kappa=2.3, \delta=0.25.
\notag
\end{align}
This yields the inverse CDF (cumulative distribution function) in Fig.~\ref{fig:construction}(b) from which the boundary decision threshold, $\tau$, is computed by (\ref{eq:boundary-decision-threshold}) and applied to the GP regression results to estimate the position of the boundary.
\begin{align}
\tau&=w\max_{(x,y)\in S_1}f(x,y) + (1-w)\min_{(x,y)\in S_2}f(x,y)\label{eq:boundary-decision-threshold}\\
w&=\frac{w_1}{w_1+w_2},\quad w_i=\frac{1}{\vert S_i\vert}\sum_{(x,y)\in S_i}f(x,y).\notag
\end{align}

Following the conventions established in Sec.~\ref{sec:observ-model}, (\ref{eq:ex-synthesis-xy-pts})--(\ref{eq:ex-synthesis-xy-blocks}) specify the centroids (x coordinates $\vec{a}_1=[\ldots,a_{i,1},a_{i+1,1},\ldots]$, y coordinates $\vec{a}_2=[\ldots,a_{i,2},a_{i+1,2},\ldots]$) and dimensions $\vec{h}_1$ and $\vec{h}_2$ of the input data used in examples 1 and 2.
\begin{align}
\begin{rcases}
a_{i,1}&=\begin{cases}0.5 & 1\le i \le 11\\5.8 & 12\le i \le 22\end{cases}\\
a_{i,2}&=\begin{cases}1+0.1(i-1) & 1\le i \le 11\\1.6+0.1(i-12) & 12\le i \le 22\end{cases}\\
h_{i,1}&=0\text{ and }h_{i,2}=0.1\qquad 1\le i\le 22
\end{rcases}
\text{\footnotesize Point measurements (Figs.~\ref{fig:motivation1} and \ref{fig:motivation2})}\label{eq:ex-synthesis-xy-pts}\\
\begin{rcases}
a_{23,1}&=1.8, a_{24:25,1}=[3,3],\\
a_{26,1}&=4, a_{27,1}=4.75\\
a_{23,2}&=0.65, a_{24:25,2}=[-0.25,0.5],\\
a_{26,2}&=0.6, a_{27,2}=2\\
h_{i,1}&=0\qquad 23\le i\le 27\\
h_{i,2}&=\begin{cases}0.5 & i=24\\1 & i\in\{23,25,26,27\}\end{cases}
\end{rcases}
\text{\footnotesize Line measurements (Fig.~\ref{fig:motivation1})}\label{eq:ex1-synthesis-xy-lines}
\end{align}
\begin{align}
\begin{rcases}
a_{23:25,1}&=[1.2,1.2,1.2], a_{26:27,1}=[2,2],\\
a_{28:29,1}&=[3,3], a_{30:31,1}=[4,4], a_{32:33,1}=[5,5]\\
a_{23:25,2}&=[0.25,0.75,1.25], a_{26:27,2}=[1.15,1.9],\\
a_{28:29,2}&=[1.2,2.2], a_{30:31,2}=[1.1,2.1], a_{32:33,2}=[1,1.5]\\
h_{i,1}&=0\qquad 23\le i\le 33\\
h_{i,2}&=\begin{cases}0.5 & i\in\{23,24,25,27,32,33\}\\1 & i\in\{26,28,29,30,31\}\end{cases}
\end{rcases}
\text{\footnotesize Line measurements (Fig.~\ref{fig:motivation2})}\label{eq:ex2-synthesis-xy-lines}\\
\begin{rcases}
\{(a_{*j,1},a_{*j,2})\}_j&=\mathcal{I}_x\otimes\mathcal{I}_y,\text{ where}\\
\mathcal{I}_x&=\{0.125 + h_*\times (x-1)\}_{1\le x\le 26}\\
\mathcal{I}_y&=\{-0.575 + h_*\times (y-1)\}_{1\le y\le 10}\\
h_{*j,1}&=h_{*j,2}\defeq h_*=0.25\quad\forall j
\end{rcases}
\text{\footnotesize Inference blocks (Figs.~\ref{fig:motivation1} and \ref{fig:motivation2})}\label{eq:ex-synthesis-xy-blocks}
\end{align}

\section{Appendix B: Research perspectives}\label{app:research-perspective}
This appendix briefly highlights some previous works and aims to put the history of data fusion within the mining industry into context. From a machine learning (ML) perspective, significant advances have been made in the application of ML techniques to ore grade estimation. One useful resource is \citet{abuntori2021evaluating} which provides a comprehensive survey and makes observations on a variety of support vector regression (SVR) and neural network (NN) based approaches; it also investigates the influence of activation functions and iterative tuning of parameters. Alternative methods such as random forest (RF), extreme gradient boost (XGBoost) and k-nearest neighbour (KNN) regression were explored in \citet{maniteja2025advancing} and found to be competitive for grade modelling in an iron ore deposit. In regard to probabilistic grade predictions, \citet{christianson2023traditional} have argued that GPs offer more economical (fewer human and compute resources), more accurate and better uncertainty quantification; noting that modern GP implementations are tailored to make the most of modern computing architectures, such as multi-core workstations and multi-node supercomputers. These authors also demonstrated that GPs can gracefully accommodate left-censoring of small measurements (due to limit of detection for trace elements) in assay data. Theoretical contributions to GP research are abound in the literature, one example is GP acceleration by reorganising observed data and using tree-structured GP approximations \citep{bui2014tree}. Related efforts include approximate inference for bayesian GP regression: in \citet{lalchand2020approximate}, the posterior over hyperparameters is approximated by Hamiltonian Monte Carlo sampling, and variational inference using a factorised Gaussian mean-field. Others have developed GP models that propagate error terms using the derivative of the predictive mean function to account for the effects of input noise through a processing pipeline \citep{johnson2019accounting}.

Despite these advances, ore grade estimation utilising various supports (such as points, intervals, areas and volumes as input data and estimation target) with a high level of efficiency and accuracy had remained a significant challenge. To address the issue of inaccuracy, practitioners often resort to approximating non-point support input data (e.g., blast holes and input block models) by point data at the expense of increased computational cost, and the output support (such as blocks in a block model) are divided into smaller sub-blocks for which point estimation is appropriate due to small variation in those volumes. The resulting estimate for each block is then calculated by combining the encompassing sub-block estimates \citep{nwaila2024spatial}. This is no longer necessary with the development of IntegralGP. Historically, there has been documented attempts at fusing blasthole with exploration assays. In \citet{boyle2011rc} for instance, two single-source estimates (one for blastholes, another for exploration data) are obtained and then merged using inverse weights based on their kriging variance. This approach incurs approximation errors as point-based kriging was used on both blastholes and exploration data. Furthermore, blastholes were filtered to exclude short holes with lengths less than 6 m and low grade samples (Fe < 50) were excluded from semi-variogram calculation so that ``the sills are lower and spatial continuity is easier to interpret.'' This means valuable data is discarded---not to mention the possibility of introducing significant estimation bias---because the modelling system at the time could not cope with inherent variability in real data. In this paper, data fusion is handled seamlessly by heteroscedastic IntegralGP (as described in Algorithm~\ref{alg:fusion}) without misgivings about the sampling configuration or noise present in the data.

\section{Appendix C: Practical considerations}\label{app:practicalities}
This appendix seeks to provide further guidance on design choices and computational tools. Technical aspects such as anisotropic kernels, hyperparameters optimisation and implementation approaches will be briefly considered.

\subsection{Isotropic vs anisotropic kernels}\label{app:techcon-anisotropic-kernel}
The covariance kernel plays a critical role in GP since it makes assumptions about the dependence between data points. Hence, the kernel choice has implications for regression performance and interpretability \citep{marcinkevivcs2023interpretable}. In \citet{soleimani2024analyzing}, it was found that anisotropic kernels (those with heterogeneous or independent length scale hyperparameters for each data dimension) outperform isotropic kernels in critical heat flux prediction. For ore grade estimation, this general finding is corroborated by \citet{jafrasteh2018comparison} and \citet{leung2024porphyry}. Some indicative figures are shown in Table~\ref{tab:aniso-vs-isotropic}. While these studies all point toward the superiority of anisotropic kernels, the improvement and application context varied. The Soleimani study uses physical parameters (such as pressure and mass flux) as input which are not compatible in scale; hence data standardisation is essential. In the separate studies conducted by Jafrasteh and Leung, spatial coordinates serve as the primary input, and far smaller differences---although still substantial improvements---were observed with anisotropic kernels in the context of grade prediction at two porphyry copper deposits. The final remark is that with anisotropic kernels, directional dependence may complicate parameter estimation and interpretation. In contrast, the effective range can be defined more intuitively for isotropic kernels based on the kriging (experimental variogram fitting) perspective.

\begin{table*}[!h]
\centering\small
\resizebox{\textwidth}{!}{
\begin{tabular}{l|p{20mm}p{12mm}p{12mm}|p{20mm}p{12mm}|p{28mm}p{12mm}p{12mm}}\hline
& \multicolumn{3}{c|}{\citet{soleimani2024analyzing}} & \multicolumn{2}{c|}{\citet{jafrasteh2018comparison}} & \multicolumn{3}{c}{\citet{leung2024porphyry}}\\
kernel type & kernel name & MAPE & RMSE & kernel name & NMSE & kernel name & $h_{JS}$ & $F$\\\hline
isotropic & Mat\'ern 3/2 & 1.1041 & 12.613 & GP-std & 0.602 & OK Mat\'ern $\nu$ & 0.3524 & 0.6234\\
anisotropic & Mat\'ern 3/2 & 0.5207 & 4.4279 & GP-SS-AK & 0.516 & GP(L)\,Mat\'ern\,3/2 & 0.2802 & 0.8231\\\hline
rel.\,change & & -52.9\% & -64.9\% & & -14.3\% & & -20.5 & +32.0\%$^\ddag$\\\hline
\multicolumn{1}{c}{} & \multicolumn{8}{p{145mm}}{\footnotesize Distortion measures: MAPE=mean absolute percentage error, RMSE=root mean squared error, NMSE=normalised mean square error, $h_{JS}$=Jensen-Shannon histogram distance. Quality measure: $F$=spatial fidelity statistic derived from variogram comparison. $^\ddag$ For the fidelity measure, a positive change indicates the anisotropic kernel has higher performance.}\\
\end{tabular}
}
\caption{Multiple studies demonstrate the superiority of anisotropic kernels over isotropic kernels.}\label{tab:aniso-vs-isotropic}
\end{table*}

\subsection{Hyperparameters optimisation}\label{app:techcon-hyperparam-optimisation}
Hyperparameters optimisation is a big topic, it is difficult to do it justice. Based on novelty and significance, one exemplary work is \citet{seifi2020modeling} where the authors have combined meta-heuristics (e.g., grasshopper optimisation, genetic and particle swarm algorithms) with neural network, adaptive neuro-fuzzy inference system (ANFIS) and SVM to create hybrid models that can reduce the RMSE in groundwater level prediction by 14\%. Another stream of research focuses on Physics-Informed Gaussian Process \citep{alvarez2013linear} which seeks to learn optimal parameters for operational equations. One example is the Ezati study which uses GP to find solution, $u(X)$, to a heat equation which contains its own model parameters $\xi$ (e.g., thermal diffusivity coefficient) and a forcing term, $f(X)$. The application in \citet{ezati2024novel} differs from modelling a geospatial phenomenon, as there is a dynamic aspect to the relationship. Both $u(X)$ and $f(X)$ are modelled using GP kernels, thus Bayesian inference requires estimation of $\xi$ and the hyperparameters for the kernels $\varphi_{uu}$, $\varphi_{ff}$ and $\varphi_{fu}$ \citep{raissi2017machine}. This can be costly due to the necessity of covariance matrix inversion and calculation of the determinant in each LML evaluation. Although the objective function is non-convex, conjugate gradient (CG) methods can be employed since the explicit form of the gradient is known.

Ezati et al. demonstrated that the initial step length and search direction used in CG methods can significantly affect the computational cost (viz., number of evaluations of the likelihood function) and whether convergence is achieved. Both aspects were improved by implementing a normalisation and restart strategy that overcomes ill-posed conditions when the search direction falls out of alignment with the direction of steepest descent. These findings relate to modelling gene cleavage in Drosophila, where parameters were estimated for a reaction-difusion PDE that describes gap gene dynamics of proteins. Currently, anecdotal evidence would suggest that learning speed and convergence are less of an issue compared with accurate domaining \citep{atalay2025domaining} within mineral resources estimation. In mining geology, stochasticity, non-stationarity and multi-modality can often be managed by partitioning the modelling space into multiple geological domains to reduce the complexity of the target distribution; as well as the size of training sets. However, future research may provide greater clarity on the benefits of improved CG optimisation algorithms. For general purpose orebody univariate grade modelling, random parameters initialisation within a reasonable search range (e.g., $l_x,l_y\in[1,100]$ for Fe and SiO\textsubscript{2}) has worked well in the authors' experience. Some optimisation and implementation strategies will be further described in the next section.

\subsection{Implementation}\label{app:techcon-implementation}
For this paper, the experiments were run on a Windows operating system using prototype code written in \textsc{Matlab}---an interpreted high-level language. The formulas for the covariance and gradients were implemented and the negative LML cost objective was wrapped inside a \code{gplmlgrad} function and passed as an argument to \code{fminunc}, a built-in nonlinear solver, which finds minimum for unconstrained multivariate functions. By default, a quasi-Newton algorithm is used, and the options \code{[LargeScale=`off', MaxIter=500, TolFun=1e-7, MaxFunEvals=500]} are applied. The production code used on large-scale mine data is written in an object-oriented programming language. It makes extensive use of \CC templates and the Factory design pattern \citep{alexandrescu2001modern} which provides an abstract interface for the proposed kernels and implements the specifics ($\varphi,\varphi',\Phi$ and $\Psi$) in concrete classes.

GP learning and inference require matrix inversion via Cholesky decomposition. As expected, efficient computation requires a linear algebra package that supports parallel execution. \textsc{Plasma} \citep{agullo2009numerical,kurzak2013multithreading} is one such library that supports multi-threading and optimised instructions for multi-core architecture. For hyperparameters optimisation, a global search followed by local search was found to be highly effective. Using the NLopt nonlinear optimisation package \citep{johnson2014nlopt}, a controlled random search with local mutation \code{nlopt::GN\_CRS2\_LM} \citep{kaelo2006some} is first attempted. This is followed by a limited memory BFGS algorithm \code{nlopt::LD\_LBFGS} \citep{liu1989limited,vlvcek2015modified} which uses conjugate direction corrections to satisfy quasi-Newton conditions. The lower and upper bounds are usually initialised to $10^{-6}$ and $10^3$, respectively, for length scales while upper-bounds for the amplitude and noise parameters are set nominally to twice the standard deviation of the target variable based on training data. The tolerance is set to $10^{-4}$, but subsequently reduced during local search. It is not uncommon for these tasks to be scheduled, ran and managed using distributed computing paradigms \citep{hwang2013distributed}. Cloud computing infrastructures provide an effective means for scaling to large datasets.

\section*{Author contributions}
Conceptualization (AC, AM), Methodology (AC, AM), Investigation (RL, AC), Formal analysis (AC, RL), Software (AC, RL), Visualization (AC, RL), Writing - Original Manuscript (RL), Writing - Review \& Editing (AC, AM), Supervision (AM).

\section*{Acknowledgements}
This work was supported by the Australian Centre for Robotics and Rio Tinto Centre for Mine Automation at the University of Sydney.

\bibliographystyle{plainnat}
\bibliography{references.bib}

\end{document}